%% file: thesis.tex
\documentclass{mythesis}
\usepackage{mythesis}
\pdfoutput=1

\makeatletter
\@ifpackageloaded{hyperref}{%
\hypersetup{%
pdftitle = {SciFi near detector for NF},
pdfsubject = {Rosen Matev's Master thesis},
pdfkeywords = {near detector, scintillating fiber, neutrino factory},
pdfauthor = {\textcopyright\ Rosen Matev}no 
}
}{}
\makeatother

\title{Scintillating fiber near detector for the Neutrino Factory}
\author{Rosen Ivanov Matev}

\begin{document}

\begin{frontmatter}
  \input{frontmatter}
\end{frontmatter}

\begin{mainmatter}
  \input{introduction}
  \input{nuescatter}
  \input{detector}

  \input{monte-carlo}
  \input{reconstruction}
  \input{analysis}
  \input{conclusions}

\end{mainmatter}

\begin{appendices}
  \input{appendices}

\end{appendices}

\begin{backmatter}
  \input{backmatter}
\end{backmatter}

\end{document}

%% file: frontmatter.tex
\titlepage[Близък детектор от сцинтилационни влакна за Неутринната фабрика]%
{Росен Иванов Матев}%
{Софийски~университет~,,Св.~Климент~Охридски'' \\ Физически факултет \\ катедра ,,Атомна физика''}%
{Научен ръководител: \\ доц. дфн Румен Ценов}%
{София, 2011 г.}%
{ДИПЛОМНА РАБОТА}
\clearpage
\clearpage
\titlepage[Scintillating fiber near detector for the Neutrino Factory]%
{Rosen Ivanov Matev}%
{University of Sofia ``St. Kliment Ohridski''\\ Faculty of Physics \\ Atomic Physics Department}%
{Supervisor: \\ Assoc. Prof. Roumen Tsenov, Dr. Sc.}%
{Sofia, 2011}%
{Master Thesis}

\begin{abstract}
\NuFact is a facility for future precision studies of neutrino oscillations.
A so called near detector is essential for reaching the aimed precision of neutrino oscillation 
analysis.
Main task of a near detector is to measure the flux of the neutrino beam. The brilliant neutrino 
source provides opportunity for precision studies of various neutrino interaction processes in a 
near detector.
We present design concept of a scintillating fiber tracker capable of measuring the neutrino flux.
The basic idea is to separate pure leptonic neutrino-electron scattering from the overwhelming 
background of neutrino-nucleon interactions.
Monte Carlo simulation of the detector and simple event reconstruction algorithms are presented.
It is shown that proper selection cuts increase signal to background ratio from
$\sim 10^{-4}$ to \unit{30-50}{\percent}.
Two methods for extraction of signal events are developed.
\end{abstract}


%

\tableofcontents
\listoffigures
\listoftables



%% file: introduction.tex
\graphicspath{{./figures/introduction/}}

\chapter{Introduction}
\label{chap:introduction}

Neutrino was postulated by W. Pauli in 1930 to explain the continuous kinetic energy spectrum of 
electrons in beta decay. He claimed that the hypothetical particle was carrying away energy, 
momentum and angular momentum, so that conservation laws are held. The direct observation of the 
neutrino was not until 1956, when F. Reines and C. Cowan conducted their experiment at the Savannah
River nuclear reactor in South Carolina. 
It was using neutrinos generated by the reactor and measured the
\begin{equation}
\APnue + \Pproton \to \Ppositron + \Pneutron
\end{equation}
interactions with their unique signature.

\section{Neutrino oscillations}
\label{sec:neutrinoOscilations}
\subsection{History}
The concept of neutrino oscillations was first stated in 1957 by B. Pontecorvo by analogy with the 
neutral kaons oscillations. The first observation of the phenomenon was made by R. Davis in the 
Homestake experiment in the late 1960s. The experiment counted the number of interactions of 
neutrinos emitted by nuclear fusion in the Solar core. Only one third of the theoretically 
predicted interaction rate was observed. Many subsequent experiments confirmed the deficit.
In 2001, The Sudbury Neutrino Observatory provided the first direct evidence of solar neutrino 
oscillation. The experiment was able to detect solar neutrinos independently of flavor as well as 
only electron neutrinos. The measured total flux of solar neutrinos agrees well with theoretical 
predictions, while electron neutrinos accounted for about \unit{35}{\percent} of the total.

\subsection{Phenomenology}
To obtain a phenomenological description of neutrino oscillations one should abandon
the Standard Model postulate of zero neutrino masses. In that case, neutrino mass eigenstates do 
not necessarily have equal eigenvalues and can in general not coincide with flavor eigenstates. 
Therefore, a flavor eigenstate of neutrino, $| \nu_{\alpha} \rangle$ ($l = e, \mu, \tau$), is in 
general expressed as superposition of mass eigenstates, $| \nu_i \rangle$ ($i = 1,2,3$):
\begin{equation}
| \nu_{\alpha} \rangle = \sum_{i} U_{\alpha i} | \nu_i \rangle ,
\end{equation}
where $U_{\alpha i}$ is an element of the Pontecorvo-Maki-Nakagawa-Sakata (PMNS) unitary matrix 
$U$. It is commonly parameterized with three mixing angles, $\theta_{12}$, $\theta_{23}$ and 
$\theta_{13}$, and one complex phase $\delta$:
\begin{equation}
U = 
\begin{pmatrix}
1 & 0 & 0 \\ 
0 & c_{23} & s_{23} \\ 
0 & -s_{23} & c_{23}
\end{pmatrix} 
\begin{pmatrix}
c_{12} & s_{12} & 0 \\ 
-s_{12} & c_{12} & 0 \\ 
0 & 0 & 1
\end{pmatrix} 
\begin{pmatrix}
c_{13} & 0 & s_{13}e^{-i \delta} \\ 
0 & 1 & 0 \\ 
-s_{13}e^{i \delta} & 0 & c_{13}
\end{pmatrix} ,
\end{equation}
where $c_{ij} = \cos \theta_{ij}$ and $s_{ij} = \sin \theta_{ij}$.
The evolution of an initial flavor eigenstate $| \nu_{\alpha} \rangle$ according to quantum 
mechanics is given by the superposition of the evolved mass eigenstates:
\begin{equation}
| \nu(t) \rangle = \sum_{i} U_{\alpha i} | \nu_i(t) \rangle
                    = \sum_{i} U_{\alpha i} e^{-i(E_i t - p_i L)} | \nu_i \rangle ,
\end{equation}
where $E_i$ and $p_i$ are the energy and momentum of $\nu_i$, and $L$ is the traveled distance.
Due to the smallness of the neutrino mass, we can make the following approximations:
\begin{equation}
&t \approx L \\
&E_i = \sqrt{p_i^2 + m_i^2} \approx p_i + \frac{m_i^2}{2 p_i} ,
\end{equation}
where the natural units system ($\hbar = c = 1$) is used.
The three mass eigenstates $\nu_i$ have the same momentum because $\nu_{\alpha}$ is produced with 
a definite momentum $p \approx E$. The probability that $\nu_{\beta}$ is observed after a distance 
$L$ is given by
\begin{equation}
P(\nu_{\alpha} \to \nu_{\beta}) 
&= | \langle \nu_{\beta} | \nu(t) \rangle |^2 =
   \left| \sum_i U_{\alpha i} U_{\beta i}^* e^{-i \frac{m_i^2 L}{2 E}} \right|^2 \nonumber \\
&= \delta_{\alpha \beta} - 4 \sum_{i>j} \mathrm{Re}(U_{\alpha i}^* U_{\alpha j} U_{\beta j}^* 
U_{\beta i})
   \sin^2\left(\frac{\Delta m_{ij}^2 L}{4E}\right) \nonumber \\
& \qquad\ \ \ \pm 2 \sum_{i>j} \mathrm{Im}(U_{\alpha i}^* U_{\alpha j} U_{\beta j}^* U_{\beta i})
   \sin\left(\frac{\Delta m_{ij}^2 L}{2E}\right) ,
\end{equation}
where $\Delta m_{ij}^2 = m_i^2 - m_j^2$ is the mass squared difference between $\nu_i$ and 
$\nu_j$, and ``+'' is for neutrinos and ``-'' is for antineutrinos. The independent square mass 
differences are only two because of the obvious relation
$\Delta m_{12}^2 + \Delta m_{23}^2 + \Delta m_{13}^2 = 0$.
Therefore, the parameters describing neutrino oscillations in general are six: 
three mixing angles, one complex phase and two squared mass differences.
Experiments have observed oscillations in two distinct regions, one at approximately 30 times 
larger $E/L$ than the other. Therefore,
one mass splitting is much smaller than the other two: $\Delta m_{12}^2 \ll \Delta m_{23}^2 
\approx \Delta m_{31}^2$. Because of this (and the smallness of $\sin \theta_{13}$), oscillations 
in the regions
$E/L \sim \Delta m_{12}^2$ and $E/L \sim \Delta m_{23}^2$ can be 
approximated by oscillation of two neutrinos with effective mixing angle $\theta$ and squared mass 
difference $\Delta m$. The oscillation probability is then given by (for $\alpha \neq \beta$)
\begin{equation}
P(\nu_{\alpha} \to \nu_{\beta}) = \sin^2 2\theta \sin^2 \left( \frac{\Delta m^2 L}{4 E} \right) .
\end{equation}

\section{Neutrino Factory}
\label{sec:nufact}
\begin{figure}
  \includegraphics[width=\mediumfigwidth]{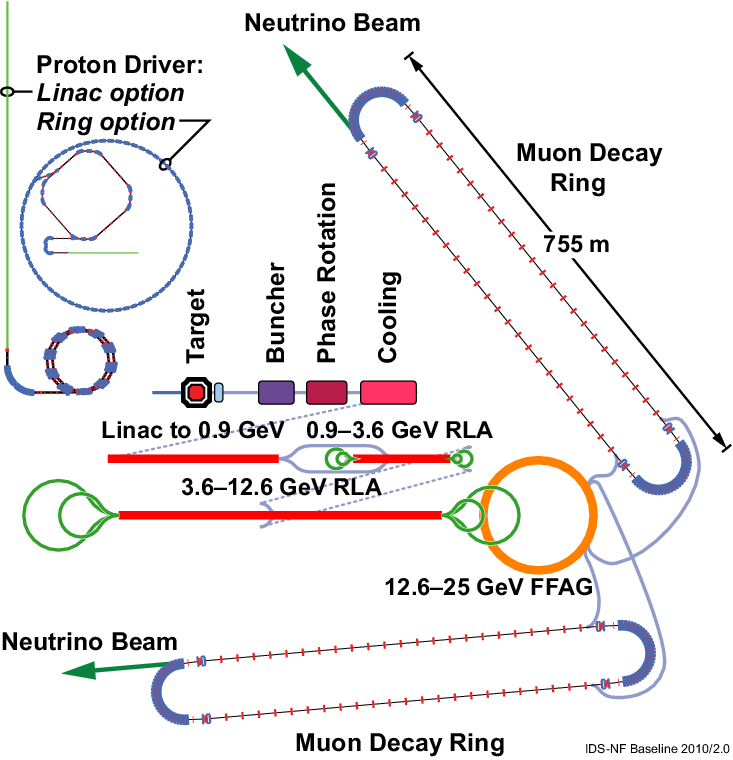}
  \caption[Baseline layout of the \NuFact accelerator complex.]%
  {Baseline layout of the \NuFact accelerator complex.}
  \label{fig:nfscheme}
\end{figure}

\NuFact is a proposed facility for precise measurement of neutrino oscillation 
parameters.
The \NuFact will create an intense neutrino beam via decays of muons stored in 
racetrack-shaped 
storage rings. The decays of negative and positive muons
\begin{equation}
\Pmuon &\to \Pelectron + \Pnum + \APnue \\
\text{and} \quad \APmuon &\to \APelectron + \APnum + \Pnue ,
\end{equation}
provide two separate neutrino beams.
The baseline layout of the accelerator complex is shown on \FigureRef{fig:nfscheme} 
\cite{Choubey:2011zz}.
Two detector locations are currently envisioned, one at ``intermediate'' 
(\unit{3000-5000}{\kilo\meter}) baseline distance and the other at 
``long''(\unit{7000-8000}{\kilo\meter}) 
baseline distance.
Specifications \cite{Choubey:2011zz} give a total of $10^{21}$ useful muon decays for both signs 
and both detector locations for $10^7 \unit{}{\second}$%
\footnote{When ``year'' or ``annual'' is used further on, it refers to $10^7\unit{}{\second}$} 
useful time.

\subsection{Accelerator complex}
For achieving the baseline muon decays per year, a \unit{4}{MW} primary proton beam is 
needed.
The unprecedented beam power will be obtained by reusing and upgrading existing 
facilities and 
building new accelerators (linear or circular).
The next step is to generate a secondary pion beam 
from interactions of the proton beam with a target. In order to avoid reinteractions of 
pions in 
the target, it has to be long and thin. This geometry imposes a great technical 
challenge due to 
the huge energy density created in the target. The solution being considered is a 
liquid jet 
mercury target \cite{McDonald:2010zz}. Generated pions will be captured by a \unit{20}{\tesla} 
solenoid, which 
tapers down to \unit{1.5}{\tesla} in a distance of \unit{15}{m}. Contrary to a magnetic 
horn, the 
solenoid allows for both pion signs to be captured.
The pion beam decays in a long decay channel, 
thus generating a tertiary muon beam with a large energy spread. The muons are bunched 
and 
phase-rotated in a sequence of RF cavities so that a train of bunches with equal 
energies is 
formed.
Following is a cooling channel, which reduces the transverse emittance of the muon 
bunches so that more muons fit in the acceptance of the following accelerators. Traditional 
techniques, such as stochastic, electron and laser cooling, are not applicable at the 
\NuFact since they take in the order of seconds to minutes to cool a beam. Muons with a rest frame
lifetime of \unit{2.2}{\micro\second} need to be cooled with a method that is fast enough. A 
novel approach is considered called ionization cooling. The idea behind the method is to use 
ionization losses to reduce emittance in both longitudinal and transverse directions and then 
accelerate to restore the initial momentum. A dedicated experiment called MICE 
\cite{Karadzhov:2010zz} to prove the feasibility of the concept is in operation.
Acceleration is divided into three stages, each chosen to use the most efficient 
acceleration 
method at its energy range. First, the muon beam is accelerated to \unit{0.9}{\GeV} 
energy by a 
linear accelerator. This is followed by two recirculating linear accelerators (RLA), 
which 
increase the energy of the beam to \unit{3.6}{\GeV} and \unit{12.6}{\GeV}, 
respectively. Finally 
the beam is accelerated in a fixed-field alternating-gradient (FFAG) accelerator. In 
both RLA and 
FFAG accelerators the beam makes many passes through the RF cavities, thus making them 
cost-effective solutions.
The beam with an energy of \unit{25}{\GeV} is injected in two racetrack-shaped decay 
rings. They 
allow simultaneous storage of both muon signs. Each decay ring points toward one of the 
far 
detector locations.

\subsection{Physics}
The \NuFact is the most promising future facility for studying of neutrino oscillation 
parameters. It 
has the following main goals:
\begin{itemize}
  \item accurately measure $\sin^2(2\theta_{13})$ if it is relatively large, and have the 
potential to test very small values otherwise.%
\footnote{Recently, the T2K experiment has shown indication that $\sin^2(2\theta_{13})$ 
is large \cite{Oyama:2011pt}. Best fits for normal and inverse hierarchies are $0.11$ and $0.14$, 
respectively. Statistical significance of the result is $2.5 \sigma$.}
  \item determine the hierarchy of neutrino mass eigenstates;
  \item measure the value of the CP violating phase $\delta$.
\end{itemize}
The advantages of the \NuFact compared to other present and future facilities are the 
intense flux, the ability to produce beams of both \Pnum (\APnum) and \APnue (\Pnue), the 
cleanliness of the beams. Twelve oscillation channels are available for studies. The six channels 
available in the (\Pnum, \APnue)-beam are shown in \TableRef{tab:channels}. The other six 
are their charge conjugates.
\begin{table}[bht]
  \caption[Oscillation channels for the \NuFact beam.]%
  {Oscillation channels for a (\Pnum, \APnue)-beam (\Pmuon decay mode). Oscillation channels in 
the other decay mode are charge conjugates of these.}
  \vspace{2ex}
  \begin{tabular}{ll}
    Channel              &  \\
    \midrule
    \Pnum \to \Pnue   & appearance ``Platinum channel'' \\
    \Pnum \to \Pnum   & disappearance \\
    \Pnum \to \Pnut   & appearance \\
    \APnue \to \APnue & disappearance \\
    \APnue \to \APnum & appearance ``Golden channel'' \\
    \APnue \to \APnut & appearance ``Silver channel'' \\
  \end{tabular}
  \label{tab:channels}
\end{table}

The Golden channel $\Pnue \to \Pnum$ has a unique experimental signature of a muon with sign 
opposite to the muon signs in the decay ring. The channel gives opportunity for measuring both
$\theta_{13}$ and $\delta$ parameters. However, there is an intrinsic degeneracy that continuously 
many solutions to the equation $P_{\alpha \beta}(\theta_{13}, \delta) = P$ exist. By studying both 
neutrino and anti-neutrino oscillations the continuous degeneracy can be eliminated. However, 
``intrinsic clone'' degenerate solutions still remain. They arise from the lack of knowledge of 
the mass hierarchy and the octant where $\theta_{13}$ is located. These degeneracies can be 
resolved at the \NuFact by studying different oscillation channels at two different baselines.

\subsection{Near detector}
\label{sec:generalRequirements}
A future neutrino facility will need near detectors in order to perform oscillation measurements 
with the required sensitivity. The near detector measurements that are essentialfor the neutrino 
oscillation analysis are \cite{Choubey:2011zz}:
\begin{itemize}
  \item measurement of neutrino flux;
  \item measurement of neutrino beam properties needed for the flux to be extrapolated to  the far 
  detector;
  \item measurement of charm production cross sections (charm production in far detectors is one
  of the principal backgrounds to the oscillation signal).
\end{itemize}
The brilliant neutrino beam allows for other standard neutrino physics studies, such as
measurement of cross sections, structure functions and $\sin^2 \theta_W$ at neutrino 
energies in the \unit{0-25}{\GeV} range.

%% file: nuescatter.tex
\graphicspath{{./figures/nuescatter/}}

\chapter{Neutrino-electron scattering}
\label{chap:nueScattering}

Neutrino-electron interactions involve only fundamental particles, thus they do not depend on 
unknown form-factors. Cross sections of these processes are straightforwardly calculated in the 
Standard Model framework. Any small uncertainties could come only from well measured 
Standard Model parameters. Therefore, such processes are suitable for measurement of neutrino 
beam fluxes, provided that beams are intense enough.
The cross sections given in this chapter are 
calculated in the low-energy approximation with four-fermion interaction. This approximation is 
justified for interactions with $s \ll m_{\PW}^2$. For a \NuFact beam having maximum energy of 
$E_{\Pnu} = \unit{25}{\GeV}$, the square of the invariant mass is
\begin{equation}
  s = (p_{\Pnu} + p_{\Pe})^2 = 2 E_{\Pnu} m_{\Pe} + m_{\Pe}^2 \approx \unit{0.03}{\GeV^2} .
\end{equation}
Neutrino-electron interactions relevant for a \NuFact beam are briefly described in the 
following sections.

\section{Quasi-elastic scattering}
\label{sec:quasiElastic}
The two purely leptonic interactions of neutrinos from the \NuFact beam producing a muon in final 
state are
\begin{equation}
  \label{eq:imd}
  \HepProcess{\Pnum + \Pelectron &\to \Pmuon + \Pnue}\\
  \label{eq:anh}
  \text{and} \quad
  \HepProcess{\APnue + \Pelectron &\to \Pmuon + \APnum} .
\end{equation}
The process in \EquationRef{eq:imd}, known as inverse muon decay, has an isotropic differential 
cross section in the center of mass frame. The total cross section is given by~\cite{Okun:100236}
\begin{equation}
  \label{eq:imdXsec}
  \sigma=\frac{G_F^2}{\pi}\frac{ (s - m_{\Pmu}^2)^{2} }{s} ,
\end{equation}
where $G_F$ is the Fermi coupling constant, $s$ is the square of the invariant mass of the system 
and $m_{\Pmu}$ is the mass of the muon.
The process in \EquationRef{eq:anh}, known as muon production through annihilation, has 
differential cross section in the center of mass frame given by \cite{Okun:100236}
\begin{equation}
  \label{eq:anhDiffXsec}
  \frac{d \sigma}{d \cos \theta} = -\frac{G_F^2}{2 \pi} \frac{(s - m_{\Pmu}^2)^2}{s^2}
    \parenths{E_{\Pe} + E_{\APnue} \cos \theta}
    \parenths{E_{\Pmu} + E_{\Pnum} \cos \theta}
\end{equation}
and the total cross section is
\begin{equation}
  \label{eq:anhXsec}
  \sigma = \frac{G_F^2}{\pi} \frac{(s - m_{\Pmu}^2)^2}{s^2}
    \parenths{E_{\Pe} E_{\Pmu} + \frac{1}{3} E_{\APnue} E_{\Pnum}} .
\end{equation}
The neutrino energy threshold (for electrons at rest) for both processes is
\begin{equation}
  \label{eq:threshold}
  E_{\Pnu} \ge \frac{m_{\Pmu}^2 - m_{\Pe}^2}{2 m_{\Pe}} \approx \unit{10.9}{\GeV} .
\end{equation}
The two quasi-elastic neutrino-electron scattering processes will be referred to with \IMD.

\section{Elastic scattering}
There are two pure neutral current reactions of interest:
\begin{equation}
  \label{eq:esnum}
  \Pnum + \Pelectron &\to \Pnum + \Pelectron \\
  \label{eq:esanum}
  \text{and} \quad 
  \APnum + \Pelectron &\to \APnum + \Pelectron .
\end{equation}
Their respective cross sections are \cite{Marciano:2003eq}
\begin{equation}
  \sigma(\Pnum \Pe \to \Pnum \Pe) &= \frac{G_F^2 m_{\Pe} E_{\Pnu}}{2 \pi}
    \parenths{1 - 4 \sin^2 \theta_W + \frac{16}{3} \sin^4 \theta_W} \quad \text{and} \\
  \sigma(\APnum \Pe \to \APnum \Pe) &= \frac{G_F^2 m_{\Pe} E_{\Pnu}}{2 \pi}
    \parenths{\frac{1}{3} - \frac{4}{3} \sin^2 \theta_W + \frac{16}{3} \sin^4 \theta_W} ,
\end{equation}
where $\theta_W$ is the Weinberg angle and $E_{\Pnu} \gg m_{\Pe}$ approximation was made.
The other processes of interest result from both \PWpm and \PZzero-boson exchange:
\begin{equation}
  \label{eq:esnue}
  \Pnue + \Pelectron &\to \Pnue + \Pelectron \\
  \label{eq:esanue}
  \text{and} \quad
  \APnue + \Pelectron &\to \APnue + \Pelectron .
\end{equation}
Their respective cross sections are \cite{Marciano:2003eq}
\begin{equation}
  \sigma(\Pnue \Pe \to \Pnue \Pe) &= \frac{G_F^2 m_{\Pe} E_{\Pnu}}{2 \pi}
    \parenths{1 + 4 \sin^2 \theta_W + \frac{16}{3} \sin^4 \theta_W} \quad \text{and} \\
  \sigma(\APnue \Pe \to \APnue \Pe) &= \frac{G_F^2 m_{\Pe} E_{\Pnu}}{2 \pi}
    \parenths{\frac{1}{3} + \frac{4}{3} \sin^2 \theta_W + \frac{16}{3} \sin^4 \theta_W} .
\end{equation}
The elastic neutrino-electron scattering processes will be referred to with \ES.

\section{Experimental signatures}
\label{sec:experimentalSignatures}
\begin{figure}
  \includegraphics[width=\mediumfigwidth]{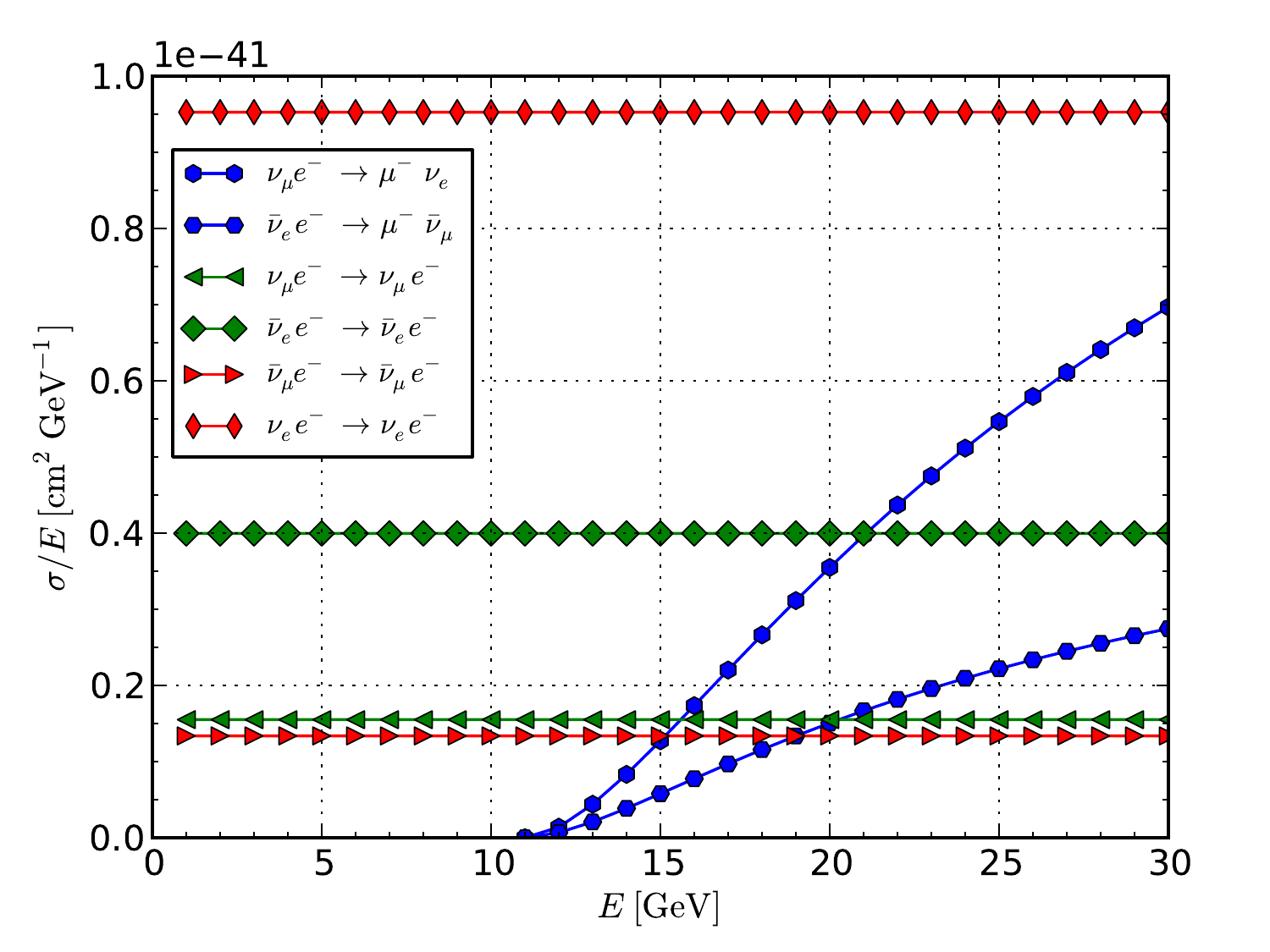}
  \caption[Neutrino-electron interactions cross sections.]{Total cross section divided by 
    neutrino energy for the leptonic interactions in a 
    \NuFact beam. Blue and green show processes which occur with \Pmuon in the storage ring, while 
    red shows processes which occur with \APmuon in storage ring. Hexagon markers show CC 
    interactions, triangle markers show NC interactions and rhombus markers show mixed CC+NC 
    interactions. The threshold for quasi-elastic processes (blue) is evident at $\sim     
    \unit{11}{GeV}$. }
  \label{fig:xsections}
\end{figure}

\begin{figure}
  \includegraphics[width=\mediumfigwidth]{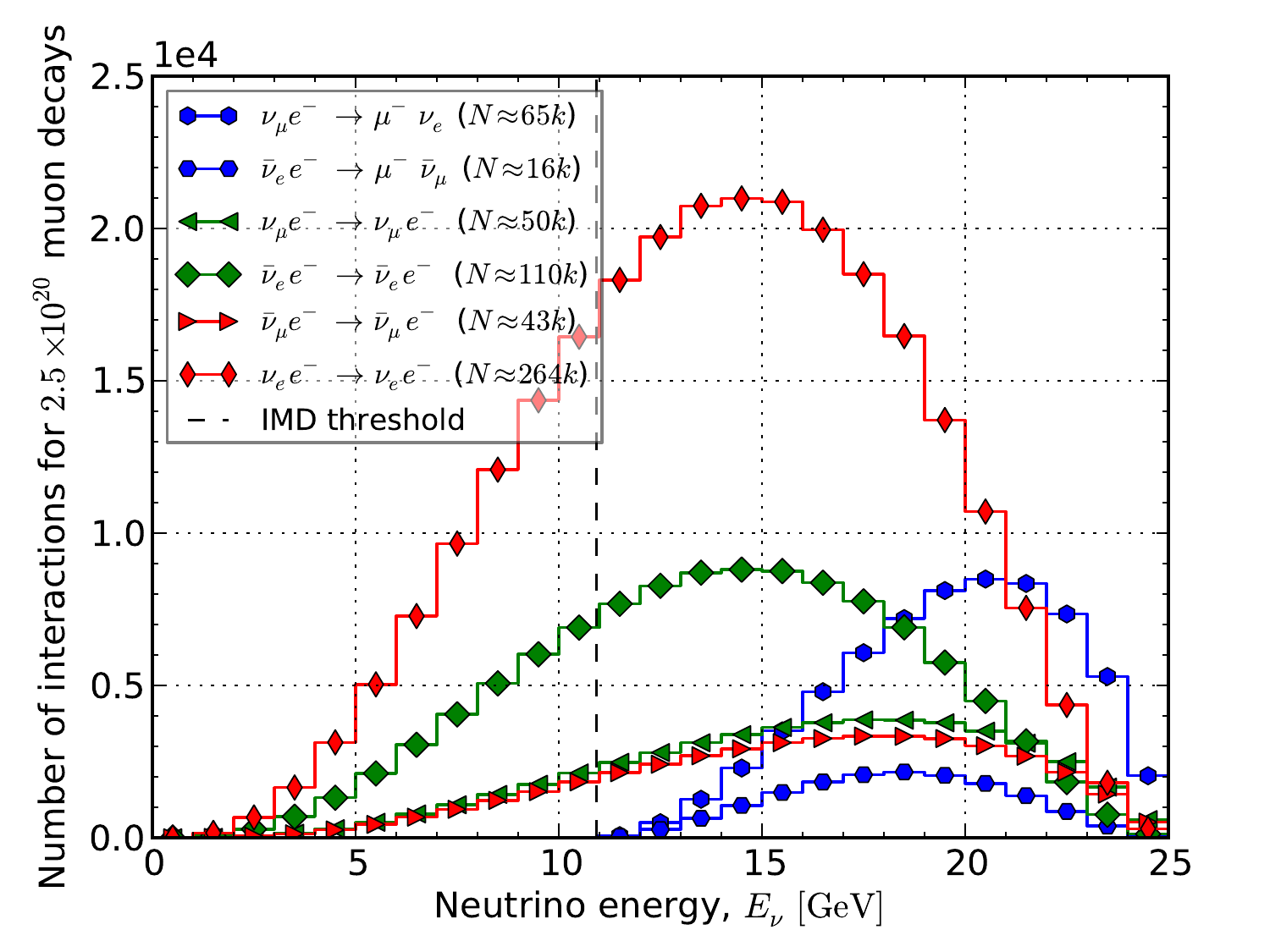}
  \caption[Neutrino-electron interactions rate.]{Number of neutrino-electron 
    interactions for a nominal year of \NuFact operation. Rates are calculated for 
    \unit{2.7}{\ton} 
    detector with \unit{1.5 \times 1.5}{\meter^2} frontal cross section and average $Z/A \approx 
    0.54$. Detector is placed \unit{100}{\meter} after the straight section end. Dashed vertical 
    line indicates threshold for quasi-elastic scattering.}
    \label{fig:eventRate}
  \end{figure}

Total cross sections for neutrino-electron interactions as a function of neutrino energy are given 
on \FigureRef{fig:xsections}. 
Despite their smallness, a couple of tons detector placed \unit{100}{\meter} after 
the straight section end can provide sufficient interaction rate - \FigureRef{fig:eventRate}. 
However, inclusive charged current (CC) and neutral current (NC) neutrino interactions with nuclei
\begin{equation}
  \label{eq:ccbgr}
  \Pnulepton + N &\to \Plepton + X \\
  \label{eq:ncbgr}
  \text{and} \quad
  \Pnulepton + N &\to \Pnulepton + X
\end{equation}
are a few orders of magnitude more frequent. For example, muon neutrino and anti-neutrino CC total 
cross sections in DIS regime are \cite{Nakamura:2010:nuxsec}
$\sigma / E_{\Pnum} \approx 0.68 \times 10^{-38} \unit{}{\cm^2 / \GeV}$ and
$\sigma / E_{\APnum} \approx 0.33 \times 10^{-38} \unit{}{\cm^2 / \GeV}$ respectively.
At \unit{15}{\GeV} the muon neutrino CC total cross section is
$\sim 1 \times 10^{-37} \unit{}{\cm^2}$,
compared to
$\sim 2 \times 10^{-41} \unit{}{\cm^2}$
for inverse muon decay $\Pnum \Pelectron \to \Pmuon \Pnue$.
An obvious distinction between purely leptonic processes and processes 
(\ref{eq:ccbgr}) and (\ref{eq:ncbgr}) is the lack of hadronic system X in the former. Measurement 
of the recoil energy of the hadronic system can be used as good criterion for background 
suppression.
Muons from quasi-elastic neutrino-electron scattering have a distribution peaked at very forward 
direction. At the \NuFact, the polar angle of \IMD muons does not exceed \unit{5}{\mrad}. The 
angular spread comes mainly from the intrinsic 
scattering angle $\sim \unit{4}{\mrad}$ in processes (\ref{eq:imd}) and (\ref{eq:anh}), while 
neutrino beam divergence (and solid angle covered by detector) has little contribution. This 
kinematic property can be used as another event selection criterion. Polar angle 
distribution of electrons from neutrino-electron elastic scattering is ten times wider and is not 
suitable for event selection. On the other hand, the \inelast variable%
\footnote{The variable \inelast is proportional to Bjorken's $y=1 - E_{\Pl}/E_{\Pnu}$ (event 
inelasticity) for elastic and approximately proportional for quasi-elastic scattering.},
provides good separation between signal and background for all neutrino-electron scattering 
processes.
  

%% file: detector.tex
\graphicspath{{./figures/detector/}}

\chapter{Near detector}
\label{chap:detector}

\section{Measurement of neutrino-electron scattering}
\label{sec:nueRequirements}
To perform measurements of neutrino-electron scattering, the detector must be able to distinguish
between the leptonic interactions and inclusive CC and NC neutrino interactions (\ref{eq:ccbgr}) 
and (\ref{eq:ncbgr}), which are a few orders of magnitude more frequent. To accomplish this task, 
the detector must make use of the distinctive event signatures mentioned in 
\SectionRef{sec:experimentalSignatures}. This imposes certain general requirements to the design 
of 
the near detector. It should be/have:
\begin{itemize}
  \item massive enough to provide sufficient interaction rate;
  \item tracker-like design to measure precisely the primary lepton's scattering angle;
  \item magnetic field to measure the momentum of the primary lepton;
  \item precise calorimeter to separate background events based on hadronic recoil energy;
  \item low Z material to minimize multiple scattering and electromagnetic showering.
\end{itemize}

\section{Scintillating fiber tracker}
\label{sec:sciFiTracker}
\begin{figure}
  \includegraphics[width=\hugefigwidth]{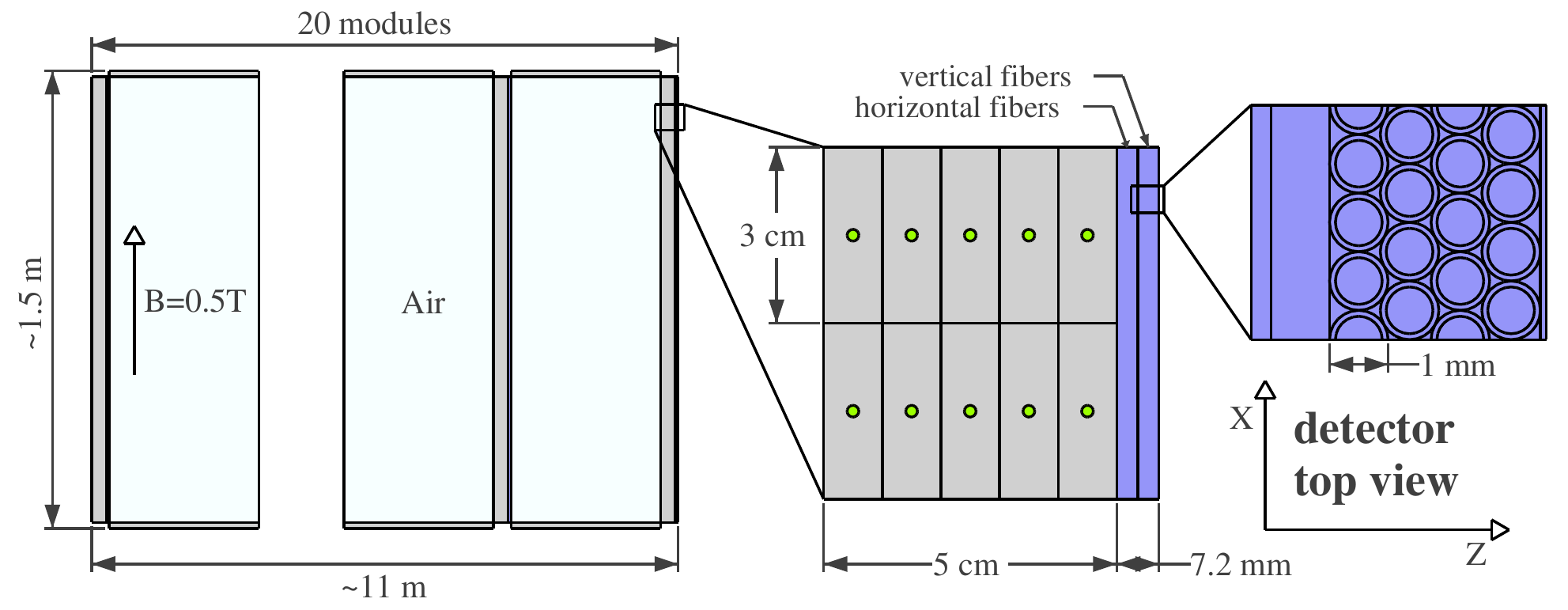}
  \caption[Schematic drawing of the detector.]%
  {Schematic drawing of the detector.}
  \label{fig:detector}
\end{figure}

A detector design which fulfills the requirements stated in \SectionRef{sec:nueRequirements} is 
proposed. The detector design is essentially a scintillating fiber tracker with an incorporated 
calorimeter.
A schematic drawing of the detector\footnote{From now on, the scintillating fiber tracker will be 
referred to as ``the detector''.} is shown on \FigureRef{fig:detector}. 
A right-handed coordinate system is adopted in which the origin is at the most upstream surface of 
the detector, the positive Z axis is aligned with the neutrino beam axis and the positive Y axis 
is pointing upwards. The detector consists of 20 square shaped modules placed perpendicular to the 
beam axis. Each module has
a calorimeter section and a tracker section (also called tracker station). Modules are positioned 
equidistantly forming gaps filled with
air. With larger distance between tracker stations, transverse displacement of hits is increased 
and thus angular resolution improved. The sides of the air gaps are covered with a layer of 
plastic scintillating bars. These layers are referred to as \emph{side slabs}.
The detector is placed in \unit{0.5}{\tesla} dipole magnetic field.
Each station consists of one layer of fibers with horizontal 
orientation and another with vertical orientation. Each layer has four planes made of 
\unit{1}{\mm} round fibers. They form a hexagonal pattern in the layer, thus minimizing dead 
volume. Calorimeter sections consist of plastic scintillating bars arranged in 5 slabs. All bars 
are oriented vertically, so that their axes are perpendicular to the magnetic field vector. Bars 
are
co-extruded with a wavelength shifting (WLS) fibers inside and have \unit{10}{\mm} by 
\unit{30}{\mm} cross-section. Both tracker fibers and WLS fibers in bars are read from both ends 
by silicon photomultipliers (SiPMs). 
Overall dimensions of the detector are $\sim \unit{1.5}{\meter} \times \unit{1.5}{\meter} \times 
\unit{11}{\meter}$. The detector mass is $\sim \unit{2.7}{\ton}$.
%

\section{Silicon photon detectors}
\begin{figure}
  \includegraphics[width=\smallfigwidth]{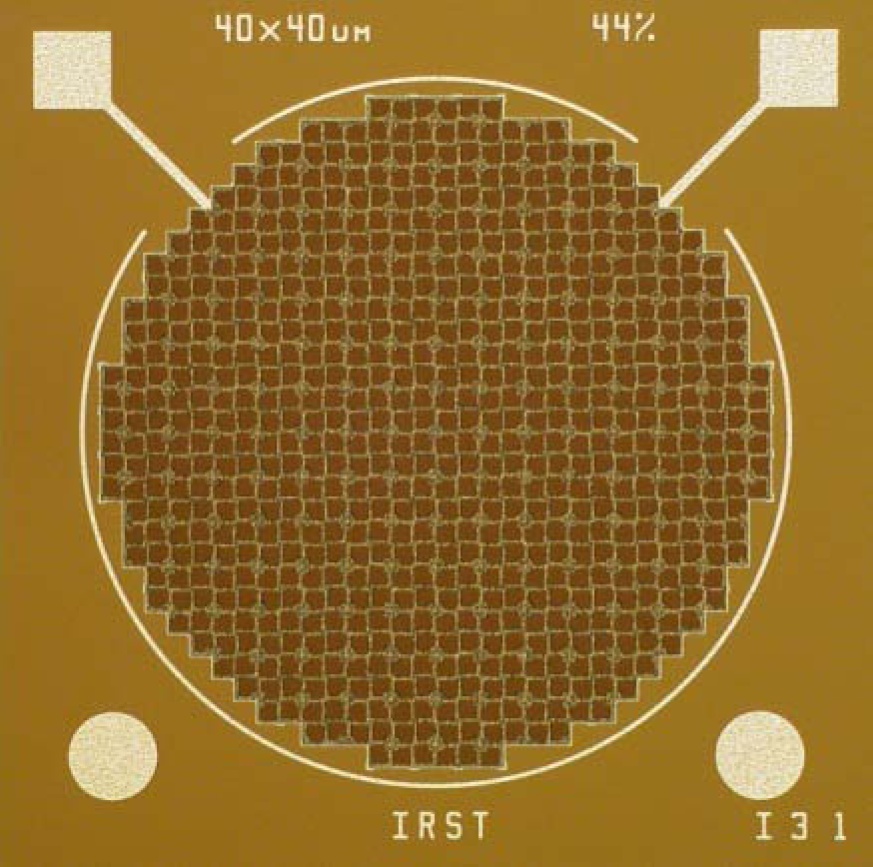}
  \caption[Photograph of SiPM.]%
  {Photograph of \unit{1.2}{\mm} diameter SiPM manufactured in FBK, Trento, Italy.}
  \label{fig:sipm}
\end{figure}

SiPM photo detectors combine many small avalanche photodiodes operated in Geiger mode to form a 
single device (see \FigureRef{fig:sipm}). They were proposed in the 1990s 
\cite{Sadygov:1996,Bacchetta:1996dc}. 
Detailed information about SiPMs can be found in a recent review paper \cite{Bacchetta:1996dc}.
In the past several years the technology has undergone rapid development, making it competitive 
to conventional photomultiplier tubes (PMT). Advantages of SiPMs over PMTs are numerous and 
include: independence of external magnetic fields, high photon detection efficiency, low operating 
voltage, compact size. The most important advantage is that SiPMs can be manufactured in standard 
microelectronics facilities. Furthermore, it is possible to integrate electronics into the SiPM 
itself, lowering the cost per channel. This makes silicon photomultipliers especially suitable for 
application in detectors having hundreds of thousands fibers, such as the scintillating fiber 
tracker.
A SiPM device is build from array of avalanche photodiodes (pixels). Size of pixels varies from 
\unit{20}{\micron} to \unit{100}{\micron}. Each avalanche photodiode is operated in Geiger mode, 
meaning that a primary trigger (photon, thermal electron) creates a discharge in the pixel. This 
results to a gain in the order of $10^6$. Each pixel is rearmed by a resistor which quenches the 
discharges by the voltage drop induced from the discharge current. All pixels are connected in 
parallel, thus forming an analog signal which is a sum of all pixel signals. Therefore, the SiPM 
signal is proportional to the number of primary triggers, provided that it is not saturated.
Although the SiPM acts as an analog device, in the case of few incident photons, the signal 
exhibits the discrete structure of the device, because all pixels are nearly the same.
The operating voltage is in the \unit{25-70}{\volt} range, a few volts above the diode's breakdown 
voltage.

The photon detection efficiency (PDE) of a SiPM depends on three factors:
\begin{equation}
\mathrm{PDE} = QE \times \varepsilon_{pt} \times \varepsilon_{geom} ,
\end{equation}
where $QE$ is the wavelength dependent quantum efficiency, $\varepsilon_{pt}$ is the probability 
of initiating a Geiger discharge by a primary trigger and  $\varepsilon_{geom}$ is the sensitive 
fraction of the device area. The gain of SiPMs is voltage dependent and varies across different
production sets of the same device. Moreover, gain has also a temperature dependence, which 
additionally complicates calibration procedures. Discharges induced by thermal electrons are 
called dark counts. Dark count rate is temperature dependent and has a typical value of less than 
\unit{1}{\mega\hertz} for \unit{1}{\mm^2} devices at room temperature. When single photon counting 
is not essential, imposing a threshold on the signal can eliminate most of the dark counts.

%% file: monte-carlo.tex
\graphicspath{{./figures/monte-carlo/}}

\chapter{Monte Carlo simulation}
\label{chap:MonteCarlo}

\section{Neutrino beam simulation}
\label{sec:mcBeam}
Neutrino flux at the near detector is generated by a Monte Carlo simulation of
muon decays along the straight section of the NF decay ring 
\cite{Karadzhov:2009zz,Karadzhov:2010su}. This \Section will briefly discuss the simulation.

For the decay of the negative muon, the expected decay distribution for the Standard Model values 
of Michel parameters is \cite{Nakamura:2010:mudecay}
\begin{equation}
\frac{d^2\Gamma}{dx d \cos \theta} &\sim x^2((3-2x) + P_\mu \cos \theta (1-2x)) \quad for\ \Pnum, 
\label{eq:muDecayMatrixM}
\\
\frac{d^2\Gamma}{dx d \cos \theta} &\sim x^2((1-x) + P_\mu \cos \theta (1-x)) \quad for\ \APnue,
\label{eq:muDecayMatrixE}
\end{equation}
where $x = 2E_\nu / m_\mu$, $P_\mu$ is the muon polarization and $\theta$ is the angle between the 
muon polarization vector and the direction of the neutrino. For the decay of 
the positive muon, the sign of the term containing $\cos \theta$ should be inverted.
Energy and angular ($\tan^{-1}(dx/dz)$) distributions of muons in the beam are taken to be 
Gaussian. It is assumed that the transverse distribution of muons in the beam pipe is a Dirac 
delta function. The parameters used for the beam simulation are given in 
\TableRef{tab:beamParameters}.
%
\begin{table}[bht]
  \caption[Parameters of muon beam in the storage ring.]%
  {Parameters of the beam in the muon storage ring used in the simulation.}
  \vspace{2ex}
  \begin{tabular}{ll}
  Parameter              & Value \\
  \midrule
  Muon energy central value & \unit{25}{\GeV} \\
  Muon energy RMS  & \unit{80}{\MeV} \\
  Muon angular spread & \unit{0.5}{\mrad} \\
  Muon polarization & \unit{0}{} \\
  Straight section length & \unit{600}{\meter} \\
  \end{tabular}
  \label{tab:beamParameters}
\end{table}

The simulation of each muon decay goes as follows. First, a position of the decay along the 
straight section is drawn from a uniform distribution. Then, for each neutrino, $x$ and $\cos 
\theta$ is drawn from distributions \ref{eq:muDecayMatrixM}, \ref{eq:muDecayMatrixE} using the 
acceptance-rejection method. An azimuthal angle $\phi$ is drawn from uniform distribution. Having 
$x$, $\cos \theta$ and $\phi$, the momentum of the neutrino in the center of mass (CM) frame of 
the muon is obtained. 
Energy and polar angle of the muon are drawn from the according Gaussian distributions. Combining 
with an uniform random azimuthal angle, one obtains the muon momentum vector in laboratory frame.
The neutrino momenta vectors in the CM frame are boosted to obtain the momenta in the 
laboratory frame. Electrons (positrons) from the decay are not simulated.

\begin{figure}
  \includegraphics[width=0.47\linewidth]{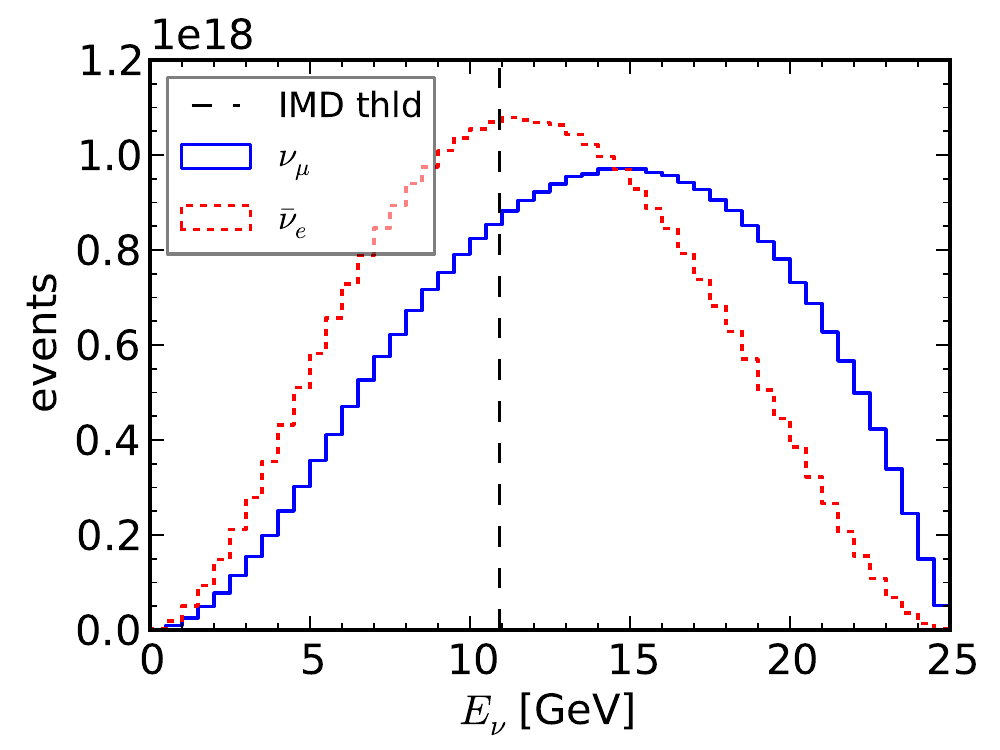}
  \includegraphics[width=0.47\linewidth]{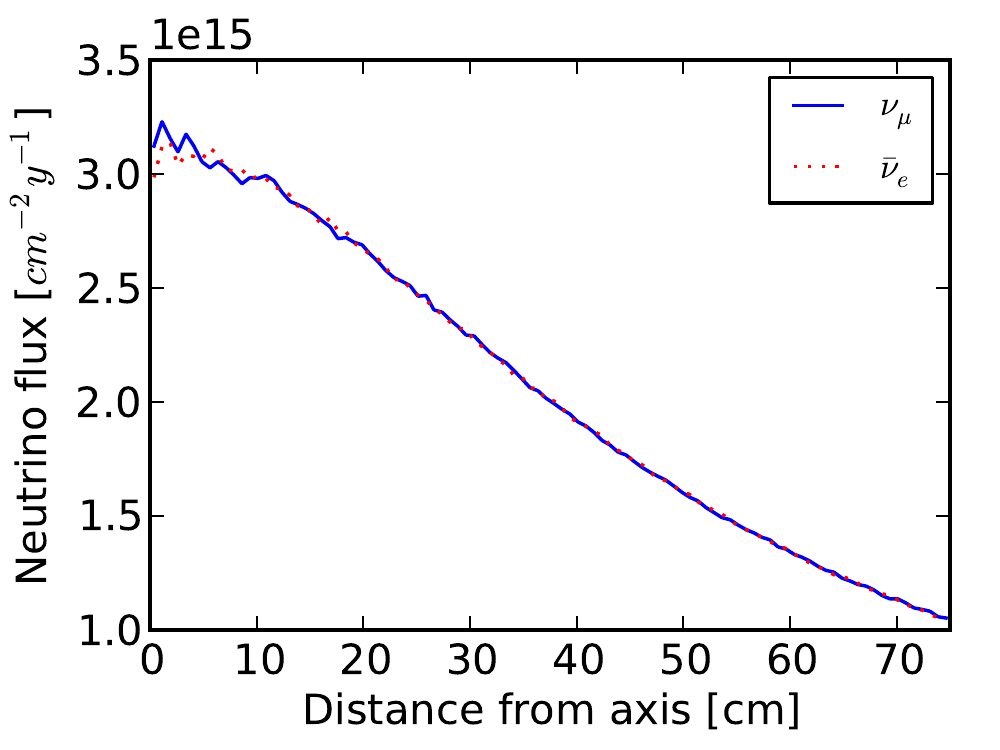}
  \caption[Neutrino flux energy and space distributions.]%
  {Energy spectrum (left) and radial 
    distribution (right) of neutrinos on plane perpendicular 
    to the straight section and \unit{100}{\meter} away from its end. The distributions are 
    normalized to the baseline specification of $2.5\times 10^{20}$ muon decays per muon charge 
    per straight section. Dashed line indicates the threshold for the leptonic processes with muon 
    in the final state.
}
  \label{fig:neutrinoFlux}
\end{figure}

The most upstream plane of the near detector is to be situated at \unit{100}{\meter} after the 
straight section end. The energy and space distributions of neutrinos at that plane is 
shown on \FigureRef{fig:neutrinoFlux}.

\section{Neutrino interactions}
\label{sec:mcInteractions}
The neutrino interactions with nuclear targets and electrons is simulated with the GENIE neutrino 
Monte Carlo generator \cite{Andreopoulos:2009rq}.
The following neutrino interaction processes are simulated by GENIE:
\begin{itemize}
  \item Quasi-elastic scattering;
  \item Elastic NC scattering;
  \item Baryon resonance production in charged and neutral current interactions;
  \item Coherent neutrino-nucleus scattering;
  \item Non-resonant deep inelastic scattering;
  \item Quasi-elastic charm production;
  \item Deep-inelastic charm production;
  \item Neutrino-electron elastic scattering; and
  \item Inverse muon decay.
\end{itemize}
The process $\APnue + \Pelectron \to \APnum + \Pmuon$ is not included in GENIE. For the simulated 
beam, the process above gives $4$ times less interactions than inverse muon decay as can be seen 
from \FigureRef{fig:eventRate}.
For this work GENIE version 2.6.2 has been used. Configuration was changed, so that short-lived 
charm particles are decayed by GENIE. Apart from this change, default values were used for all 
other parameters.

The NF neutrino flux is fed to GENIE by implementing a flux driver interface. The detector 
description used for event generation is a simplified version of the actual detector geometry. 
Each detector module is implemented as a solid box made of polystyrene. Air gaps are left as they 
are. Covering side bars are not implemented as interactions there are not in fiducial volume.
For each neutrino interaction, GENIE produces a list of intermediate and final state particles 
with their kinematic characteristics. The type of the process involved in the interaction as well 
as some kinematic variables are also provided.

\section{Detector simulation}
\label{sim:mcDetector}
\subsection{Particle transport}
For the simulation of detector response to neutrino interactions, the Geant4 
\cite{Agostinelli:2002hh, Allison:2006ve} software platform was used. Geant4 is a toolkit for 
``the simulation of 
the 
passage of particles through matter''. From the user's point of view, to simulate an event with 
Geant4 one has to define (by implementing the toolkit's interface classes):
\begin{itemize}
  \item primary particle generator;
  \item list of particles to be transported (including decay products and particles produced in 
interactions) through the detector and relevant physics processes (physics list);
  \item detector description in terms of geometric volumes and comprising materials;
  \item ``sensitive'' volumes, where ``snapshots'' of the physical interaction and state of 
particles are recorded to produce ``hits''.
\end{itemize}

\begin{figure}
  \includegraphics[width=\hugefigwidth]{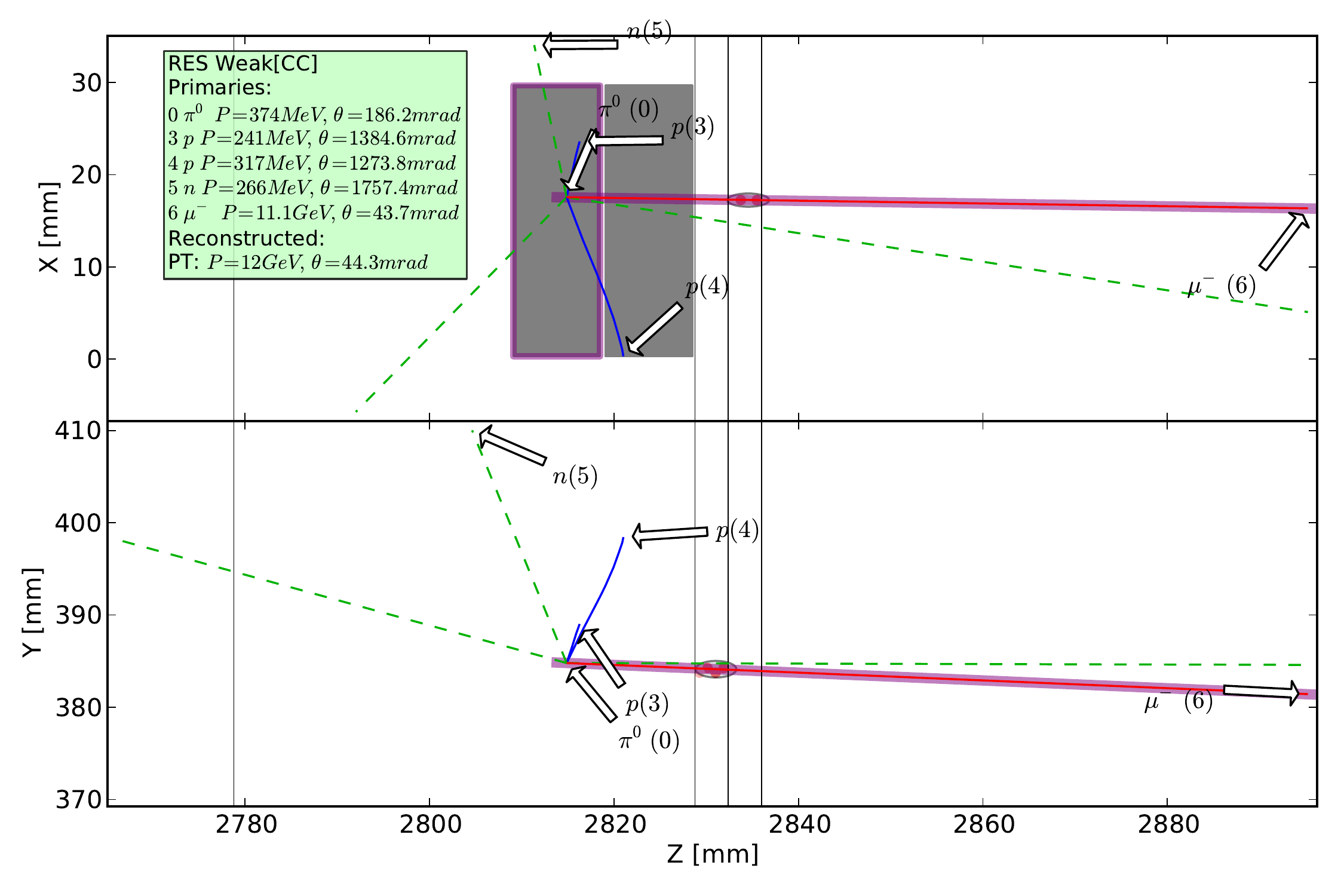}
  \caption[Example event display.]{Example event display showing a CC resonance production event.}
  \label{fig:eventDisplay}
\end{figure}

All final state particles produced by GENIE are passed to Geant4 simulation as primary particles. 
The physics list used in the simulations is ``Simple and Fast Physics List'' 
\cite{SimpleFastPhysList}.
The scintillating fiber tracker geometry was implemented in a flexible manner, so that detector 
parameters are set with a configuration file. Plastic scintillating fibers and bars are defined as
sensitive volumes. The hits produced in sensitive volumes contain: unique id of the volume, id of 
the particle, energy deposition, time and position of the interaction. True particle trajectories 
are recorded for some of the particles. Part of the information recorded in the simulation is used 
to produce electronic signals (digitization). Other data is retained for later comparison of 
reconstructed data with true data (often called ``Monte Carlo'' truth). By doing the comparison 
with MC truth, the performance of the detector design and reconstruction algorithms can be 
analyzed. A basic event display has been developed to enable visual inspection of events. Part of 
an example event display is shown on \FigureRef{fig:eventDisplay}. Apart from MC truth, the event 
display shows some reconstructed data. 

Tracking of individual particles with Geant4 is stopped when their energy is below a certain 
threshold and it is guaranteed that they will not cross a volume border. Scintillation process was 
not included in the Geant4 simulation because tracking of individual scintillating (optical) 
photons is a very time consuming task.

\subsection{Digitization}
Digitization algorithms take the output hits of the Geant4 simulation and transform them into 
electronic signals.
\subsubsection*{Fiber signal digitization}
As simulation of optical photons is not feasible in event by event basis, a toy Monte Carlo 
algorithm was developed to simulate the processes of scintillation, photon propagation into fibers 
and SiPM response. For each energy deposit in a fiber, a random number of scintillating photons is 
drawn according to the scintillation yield of the material. The scintillating photons are emitted 
isotropically, hence not all of them suffer total internal reflection from the walls of the 
fibers. The fraction of photons which do not exit the fiber is called trapping efficiency. It is 
determined from refraction indexes and fiber cross-section. For round fibers manufacturers provide 
a ``minimal'' trapping efficiency, namely the efficiency for photons propagating in a single 
central plane. However, it has been shown \cite{Papandreou:2007} that the real trapping efficiency 
is about two 
times higher than that.
A value of \unit{8}{\percent} for the trapping efficiency is used in the simulation.
The trapping efficiency and light attenuation are taken into account to produce a number of 
photons arriving at the end of the fiber. If there are multiple hits in a fiber, photons arriving 
at the fiber end are summed up. The obtained number of photons is multiplied with an optical 
coupling efficiency factor to get the number of incident photons on the SiPM surface. Multiplying 
the last number by the photon detection efficiency (PDE) one obtains the number of primary 
avalanche triggers. Dark counts are simulated by further adding a Poisson distributed number to 
the primary triggers. The final electronic signal response is parameterized by a function that 
takes into account cross-talk effect and the single pixel response distribution. Only signals with 
amplitude larger than threshold equivalent to 2.5 fired pixels are recorded. For each fiber, the 
signals from both photo detectors and an id of the fiber are stored in a structure called 
\emph{fiber digit} only if both signals are above the threshold.
It seems dark counts are not a major problem for \unit{1}{\mm} fibers and modern SiPMs, which have 
less than \unit{1}{\MHz/\mm^2} dark count rate at room temperature. With the imposed signal 
threshold and coincidence requirement the number of ``false'' fiber digits is under 
\unit{1}{\percent} of all fibers in the detector. Therefore, the probability to have two or more 
adjacent ``false'' fiber digits which mimic true particle hits is negligible.

\subsubsection*{Bar signal digitization}
As above, Geant4 simulation does not treat optical photon processes in the scintillating bars and 
the WLS fibers inside. Ideally, one can make a detailed simulation or an experiment with prototype 
bar and parameterize the signal response (wavelength spectrum) at the end of the WLS fiber. 
However, we do not have such parameterization. For this work, bar signals were naively digitized 
by smearing energy deposits with \unit{20}{\percent} Gaussian. Signals below a threshold of 
\unit{0.5}{\MeV} equivalent are discarded. For each bar, the signals 
from both photo detectors and an id of the bar are stored in a structure called \emph{bar digit}.

%% file: reconstruction.tex
\graphicspath{{./figures/reconstruction/}}

\chapter{Software reconstruction}
\label{chap:reconstruction}
Raw output from both Monte Carlo simulation and a real experiment is a set of digitized signal 
amplitudes. To understand what processes took place and what particles propagated through the 
detector, one needs to obtain measurements of relevant physical quantities characterizing them. 
The aim of reconstruction algorithms is to infer high level information about the event from the 
raw detector output.

\section{Initial processing}
The first step of reconstruction process is to obtain some basic physical quantities from the 
digital signals. First, signal from the photo detectors should be translated back 
to number of incident photons. For fibers in tracker stations this quantity is sufficient. On the 
other hand, for bars, one should further obtain an absolute measurement of the energy deposit via 
a dedicated calibration procedure. Since this is a simulation rather than real experiment, the 
above procedure is assumed perfect and the calibration step is skipped.

To simplify further the reconstruction, neighboring fired fibers are grouped into clusters. Two 
fibers are considered neighbors if their projections on the XY plane are either neighboring or 
overlapping. Depending on the fibers orientation, $(x,z)$ or $(y,z)$ position of the cluster is 
calculated by taking weighted average of the fiber positions. Weighted standard deviations from 
the mean coordinates are also calculated. A cluster object then contains list of fibers, two 
coordinates with their standard deviations and an amplitude, which is the sum of the numbers of 
incident photons on the photo detectors at the fibers. Clusters composed of only one fiber are 
discarded to eliminate false hits due to SiPM noise.

\section{Vertex reconstruction}
\label{sec:vertex}
\begin{figure}
  \includegraphics[width=\mediumfigwidth]{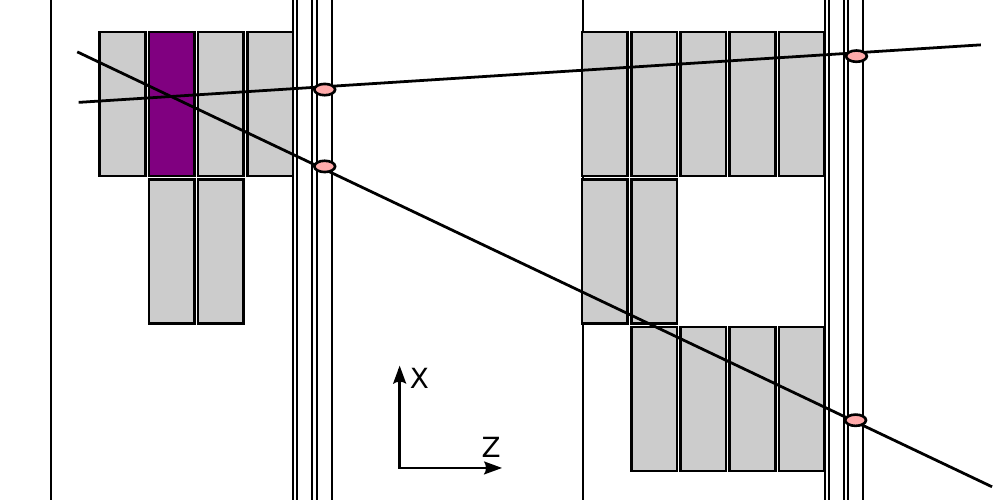}
  \caption[Vertex volume reconstruction example.] %
  {Vertex volume reconstruction example. The two 
  adjacent most upstream stations with at least one cluster each are considered. The clusters 
  (pink ellipses) with highest X coordinate are selected from both stations and a line passing 
  through their centers is taken. Another line is constructed from the clusters with lowest X 
  coordinates. The most upstream fired bar, which is between the lines (but not upstream of the 
  intersection point) is identified as vertex bar (purple box).}
  \label{fig:vertexLocator}
\end{figure}

The point where a neutrino interaction takes place is called vertex.
The scintillating fiber tracker, like a typical neutrino detector, serves as a target.
Therefore, one does not have \textit{a priori} knowledge of the precise vertex coordinates. The 
vertex reconstruction algorithm attempts to find the volume (bar or fiber), where the vertex is 
located. Only the two most upstream stations with at least one cluster are considered. 
Since particle tracks are not curved in the XZ plane, we can linearly back-project the positions 
of the clusters towards the event vertex. To do this, the clusters with highest X coordinate are 
selected from both stations and a line passing through their centers is taken. Another line is 
constructed from the clusters with lowest X coordinates. The most upstream fired bar, which is 
between the lines (but not upstream of the intersection point) is identified as vertex bar 
(\FigureRef{fig:vertexLocator}).
If there is no such bar, the cluster 
with the highest amplitude in the first upstream station is picked and its most upstream fiber is 
labeled as vertex volume. If there are no two adjacent stations with at least one cluster in each, 
the vertex cannot be reconstructed and the event is discarded. \unit{86}{\percent} of the 
events have their vertex volume reconstructed.

\begin{figure}
  \includegraphics[width=\largefigwidth]{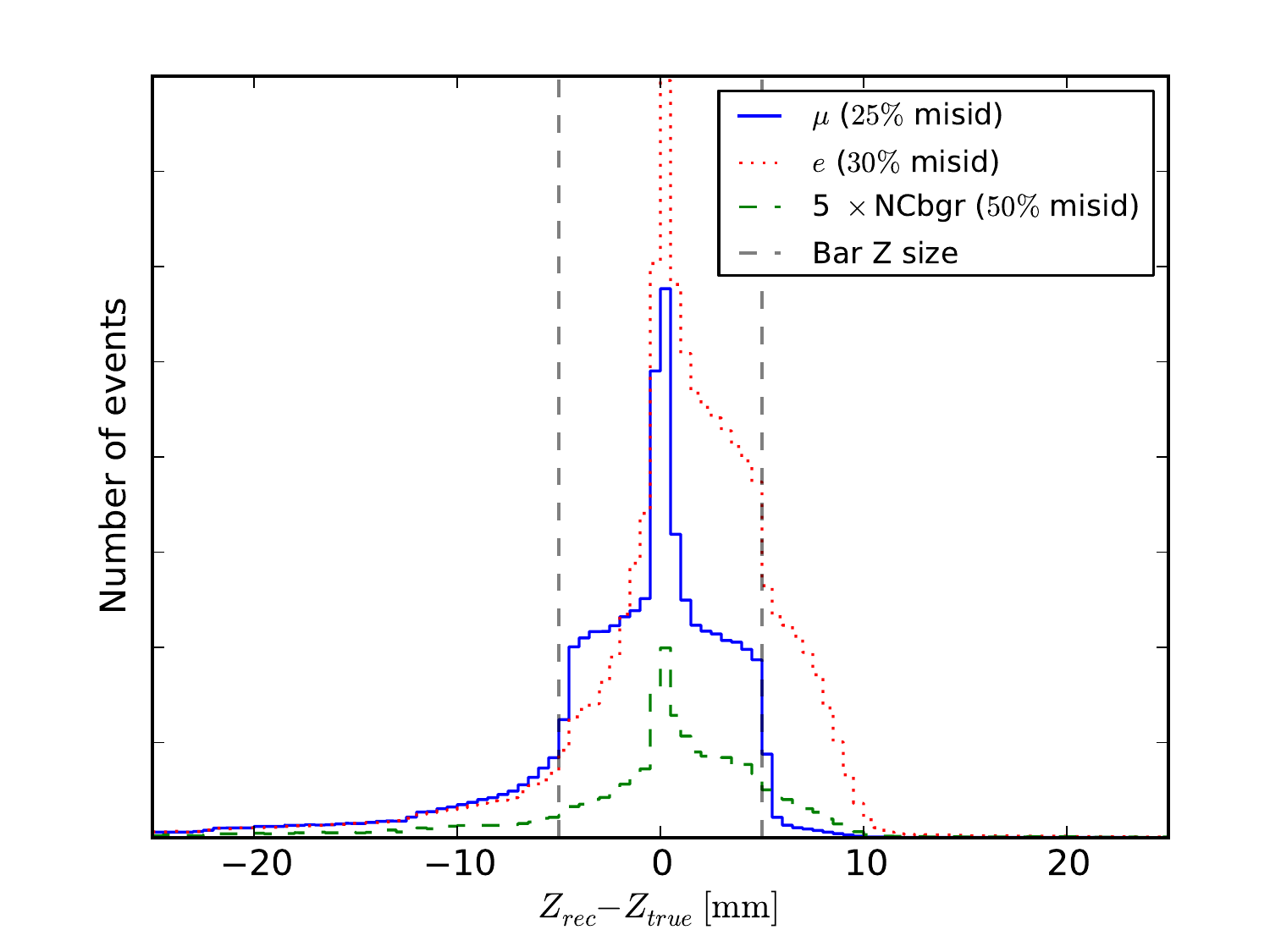}
  \caption[Vertex Z coordinate resolution.]%
  {Vertex Z coordinate resolution. Distributions of the difference 
  of reconstructed and true vertex Z coordinate are shown for CC \Pnum interactions (solid blue), 
  CC \APnue interactions (dotted red) and NC (\Pnum, \APnue) interactions (dashed green,
  $\times 5$). Vertical dashed lines indicate Z boundaries of a bar centered at 0. Uniform 
  distribution of events with
  correctly identified vertex bar is clearly visible. Events in the narrow peak centered at 0 have 
  their vertex volume correctly identified in a cluster. The shoulder on the right (and the dip 
  left of 0) for CC \APnue interactions (red) correspond to events with a reconstructed vertex bar 
  adjacent and downstream of the true vertex bar. This misidentification may be explained with 
  energy deposition in the vertex bar below the threshold and is only visible in antineutrino 
  events. 
  Misidentification probabilities for the three samples are shown in the legend.}
  \label{fig:vertexResolution}
\end{figure}

However, the vertex volume is not always correctly 
identified. The performance of the algorithm is shown on \FigureRef{fig:vertexResolution} in terms 
of difference between Z coordinate of the reconstructed volume's center and the true vertex Z 
coordinate. Difference distributions are shown for interactions of (\Pnum, \APnue)-beam.

\section{Primary track reconstruction}
\label{sim:Track}
The scintillating fiber tracker is a very fine grained detector. Because of this, reconstruction 
of the full event topology might be attempted. However, for the purpose of measuring purely 
leptonic processes, only events with a single track originating from the vertex are considered. 
The main goal of reconstruction is to identify the \emph{primary (lepton) track} and 
measure its parameters.

Only events with one cluster per orientation in the first two stations are considered. This 
selection has low efficiency, because an event with a single primary particle might still have 
more than one cluster in the either of the first stations. (Consider $\delta$-electrons or 
interactions of bremsstrahlung photons) Furthermore, as both angle and momentum of the primary 
track need to be measured, only events with hits in at least three tracker stations are 
retained. The remaining 
fraction of leptonic events after the above selections is \unit{61}{\percent} for \IMD and 
\unit{52}{\percent} for \ES. Angle and curvature of tracks is measured using only hits in the 
first three stations. In third station, if there are more than one clusters, only the cluster 
nearest to the straight line defined by clusters in the first two stations is considered. At the 
end, for the primary track we have: initial position, initial slopes ($x'_0 = 
dx/dz$ and $y'_0 = dy/dz$), initial momentum and charge.

\begin{figure}
  \includegraphics[width=0.49\textwidth]{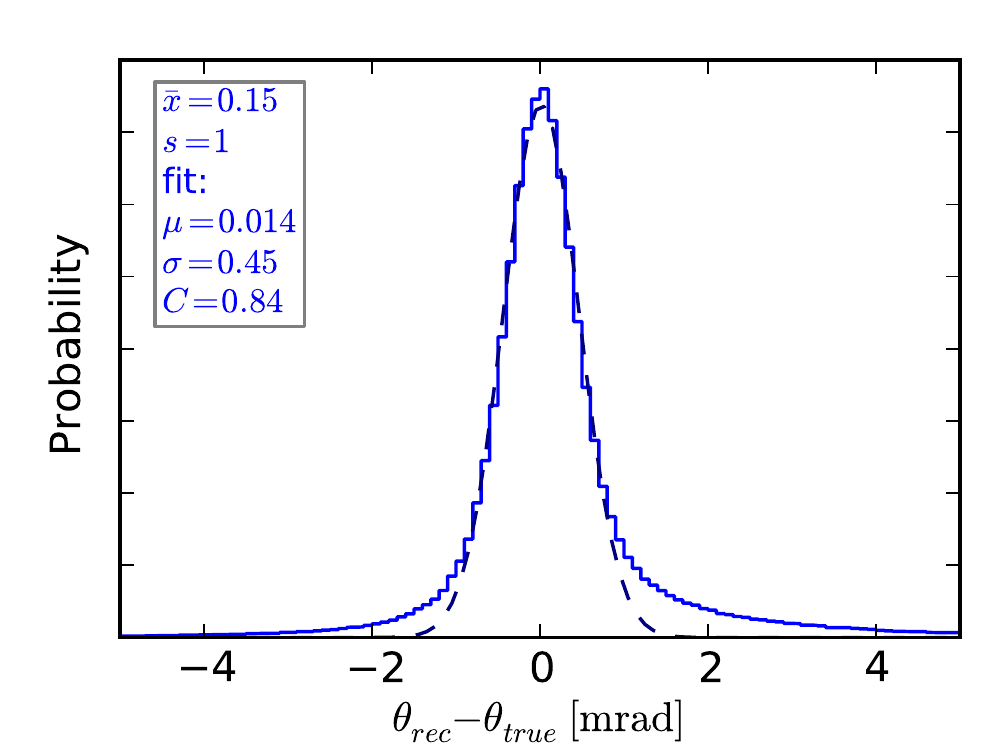}
  \includegraphics[width=0.49\textwidth]{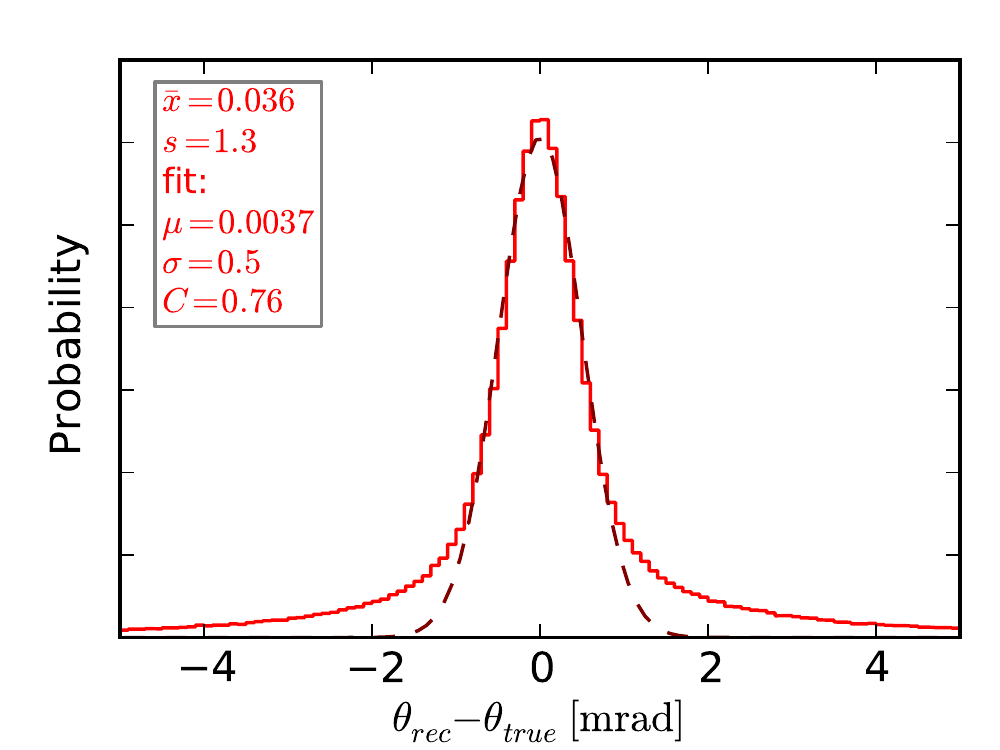}
  \caption[Angular resolution of primary track.]{Obtained angular resolution for muons (left) 
and electrons (right). Gaussian fits are shown with dashed lines. The sample mean, standard 
deviation and the fit parameters are shown in upper left corners.}
  \label{fig:angularResolution}
\end{figure}

\begin{figure}
  \includegraphics[width=0.49\textwidth]{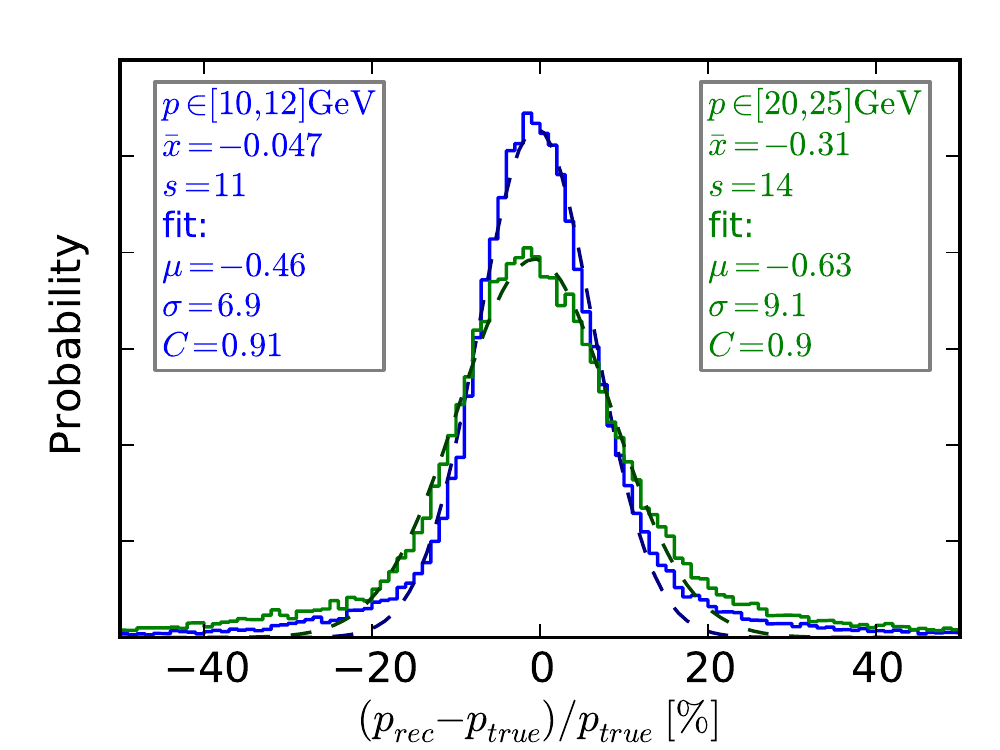}
  \includegraphics[width=0.49\textwidth]{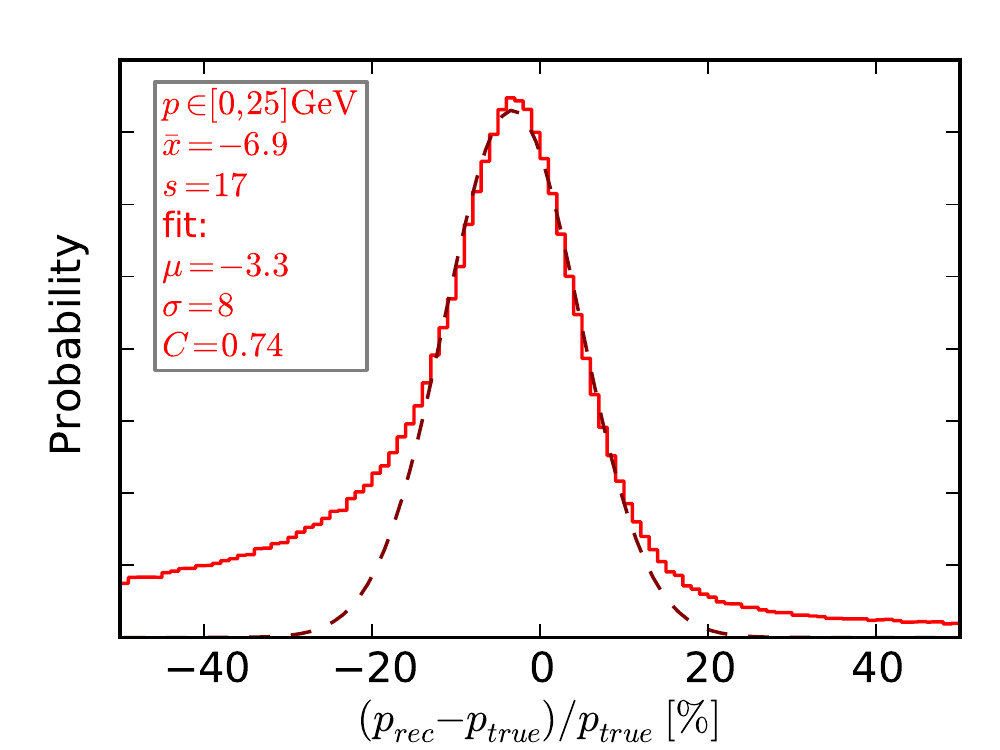}
  \caption[Momentum resolution of primary track.]{Obtained momentum resolution for muons (left) 
and electrons (right). Gaussian fits are shown with dashed lines. The sample mean, standard 
deviation and the fit parameters are shown in upper corners. For muons, distribution is shown for 
two samples of events: one with true muon momentum in the \unit{[10-12]}{\GeV} range (blue) and 
one in the \unit{[20-25]}{\GeV} range (green).}
  \label{fig:momentumResolution}
\end{figure}

To asses the performance of the primary track reconstruction algorithm, the reconstructed track 
initial angles and momenta are compared to MC truth. Distribution of difference between 
reconstructed initial angle of the primary track $\theta_{rec}$ and true initial angle of the 
primary lepton $\theta_{true}$ is shown on \FigureRef{fig:angularResolution}. From the Gaussian 
fit to the distribution, it is seen that the reconstructed angle is unbiased and the 
resolution ($\sigma$ parameter of the fit) is $\sim \unit{0.5}{\mrad}$ for both muons and 
electrons. Distribution of relative difference  between reconstructed initial momentum of the 
primary track and true initial momentum of the primary lepton
\begin{equation}
\Delta = \frac{p_{rec}-p_{true}}{p_{true}} ,
\end{equation}
is shown on \FigureRef{fig:momentumResolution}. For muons, the Gaussian fits show that the 
reconstructed momentum have only small bias (due to energy loss) and resolution is going up to 
$\sim \unit{9}{\percent}$ for the highest energy muons. For electrons, the distribution is biased 
towards the negative values with a heavy negative tail. The reason for this underestimation of 
electrons' initial momenta is that they loose momentum due to bremsstrahlung and ionization. If 
energy loss is taken into account with Kalman filter fitting, for instance, bias can be reduced. 
Kalman filter fitting \cite{Kalman:1960} is a powerful method, which can improve angular and 
momentum resolutions. This is true especially if full advantage of pattern recognition abilities 
of the method are used.

\section{Spatial distribution of energy deposits}
\label{sec:meandev}
Having this rather simple event reconstruction, a backward going particle will cause wrong 
reconstruction of vertex position. Moreover, the backward going track will be wrongly identified 
as a primary lepton track. Some background events have particles which do not cross 
the first tracker station after the vertex and do not deposit enough energy in the vertex volume. 
In order to select events with a single primary particle one might exploit the spatial symmetry 
of energy deposits relative to the track. A variable which has good signal/background event 
separation power is the following.
Firstly, we calculate the energy deposit weighted average position of the hits in the bars in the 
i-th slab:
\begin{equation}
  x_i^{obs} = \frac{\sum_{j} x_{ij} \Delta E_{ij}}{\sum_{j} \Delta E_{ij}} ,
\end{equation}
where $j$ denotes the bar in the slab, $x_{ij}$ is the X coordinate of the center of the bar and 
$\Delta E_{ij}$ is the energy deposit in the bar. Secondly, the expected position of the 
primary track at the i-th slab is calculated:
\begin{equation}
  x_i^{exp} = x_0 + x'_0 (z_i - z_0) ,
\end{equation}
where $z_i$ is the Z coordinate of the i-th slab's center, $x_0$ and $z_0$ are the primary track's 
initial coordinates and $x'_0 = dx/dz$ is the primary track's initial slope.
Finally, the variable is
\begin{equation}
  x_{dev} =  \frac{1}{n} \sum_{i} |x_i^{obs} - x_i^{exp}| ,
\end{equation}
where the summation is over all fired slabs. Such variable is expected to have low values for 
events with a single muon, medium values for electron showers and high ones for high multiplicity 
asymmetric events.

Overall reconstruction efficiency (remaining signal events) is \unit{53}{\percent} for \IMD and 
\unit{45}{\percent} for \ES. Inefficiencies come from inability to reconstruct the vertex, 
selection of events based on the number of clusters in the two most upstream fired stations and 
inability to reconstruct primary track's initial angle and momentum.

%% file: analysis.tex
\graphicspath{{./figures/analysis/}}

\chapter{Data analysis}
\label{chap:analysis}

\section{Neutrino-electron scattering event selection}
\label{sec:eventSelection}
The \NuFact beam is composed of \Pnum and \APnue neutrinos in \Pmuon decay mode and from \APnum 
and \Pnue neutrinos in \APmuon decay mode. Since we can distinguish interactions only by their 
final state, we should group neutrino-electron interactions as follows:
processes (\ref{eq:imd}) and (\ref{eq:anh}) in the \emph{\IMD} group; 
processes (\ref{eq:esnum}) and (\ref{eq:esanue}) in the \emph{\ESm} group;
processes (\ref{eq:esanum}) and (\ref{eq:esnue}) in the \emph{\ESp} group.
Sometimes \emph{\ES} will be used when referring to both \ESm and \ESp.
Three neutrino-electron event samples are selected from all events corresponding to the three 
groups. 
\IMD and \ESm samples are selected in \Pmuon decay mode, while \ESp sample is selected in \APmuon 
decay mode.

\subsection{Calorimetric vertex selections}
\begin{figure}
  \includegraphics[width=\largefigwidth]{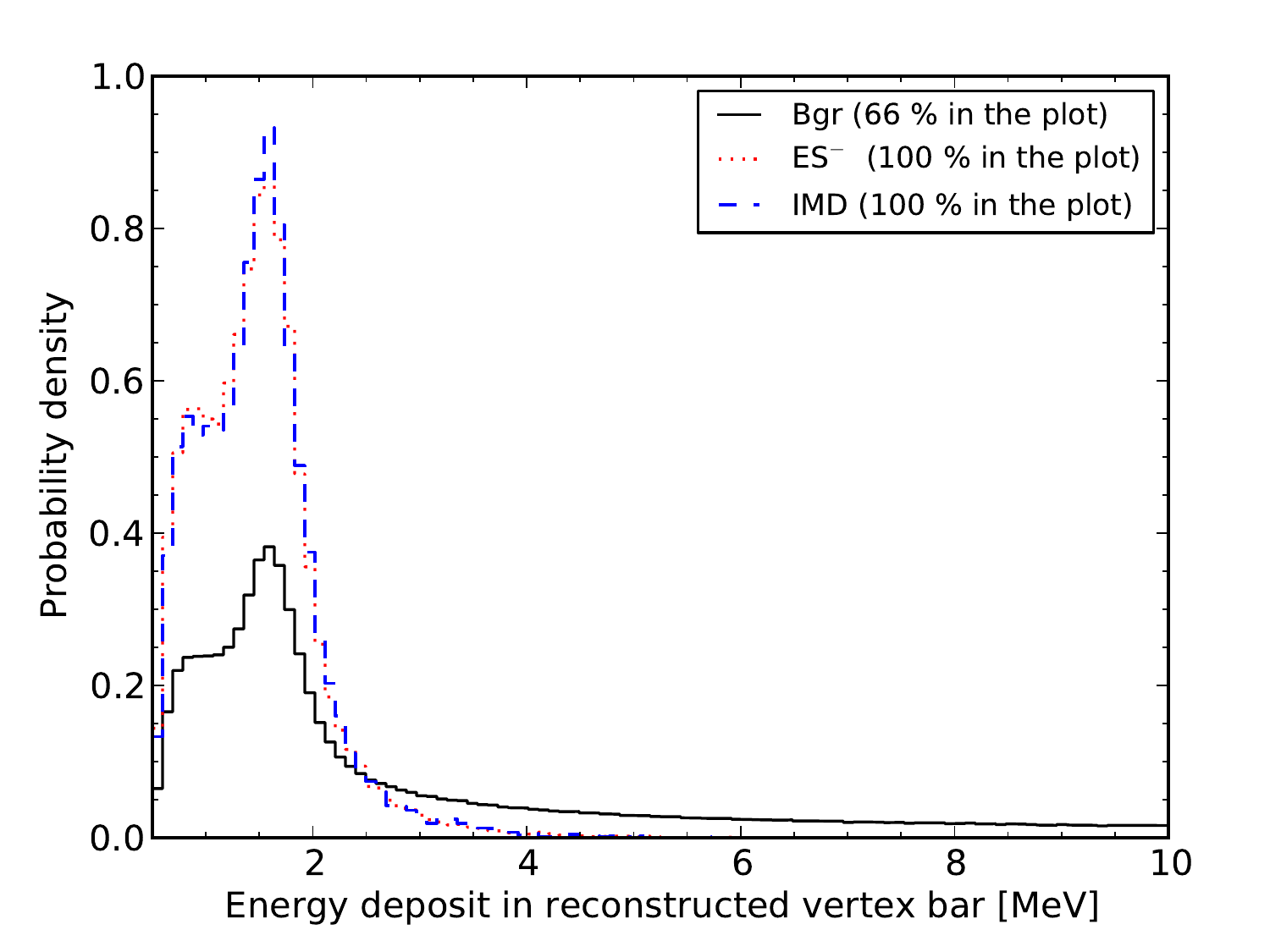}
  \caption[Distributions of energy deposit in vertex bar.]%
  {Distributions of energy deposit in vertex bar for \IMD (blue), \ESm (red) and background 
  (black) events in \Pmuon mode. The fraction of events contained in the plot is indicated in the 
  legend. All distributions are normalized to a unit area. This plot shows only events in which 
the 
  reconstructed vertex volume is a bar.}
  \label{fig:bgrVertex}
\end{figure}

\begin{figure}
  \includegraphics[width=\largefigwidth]{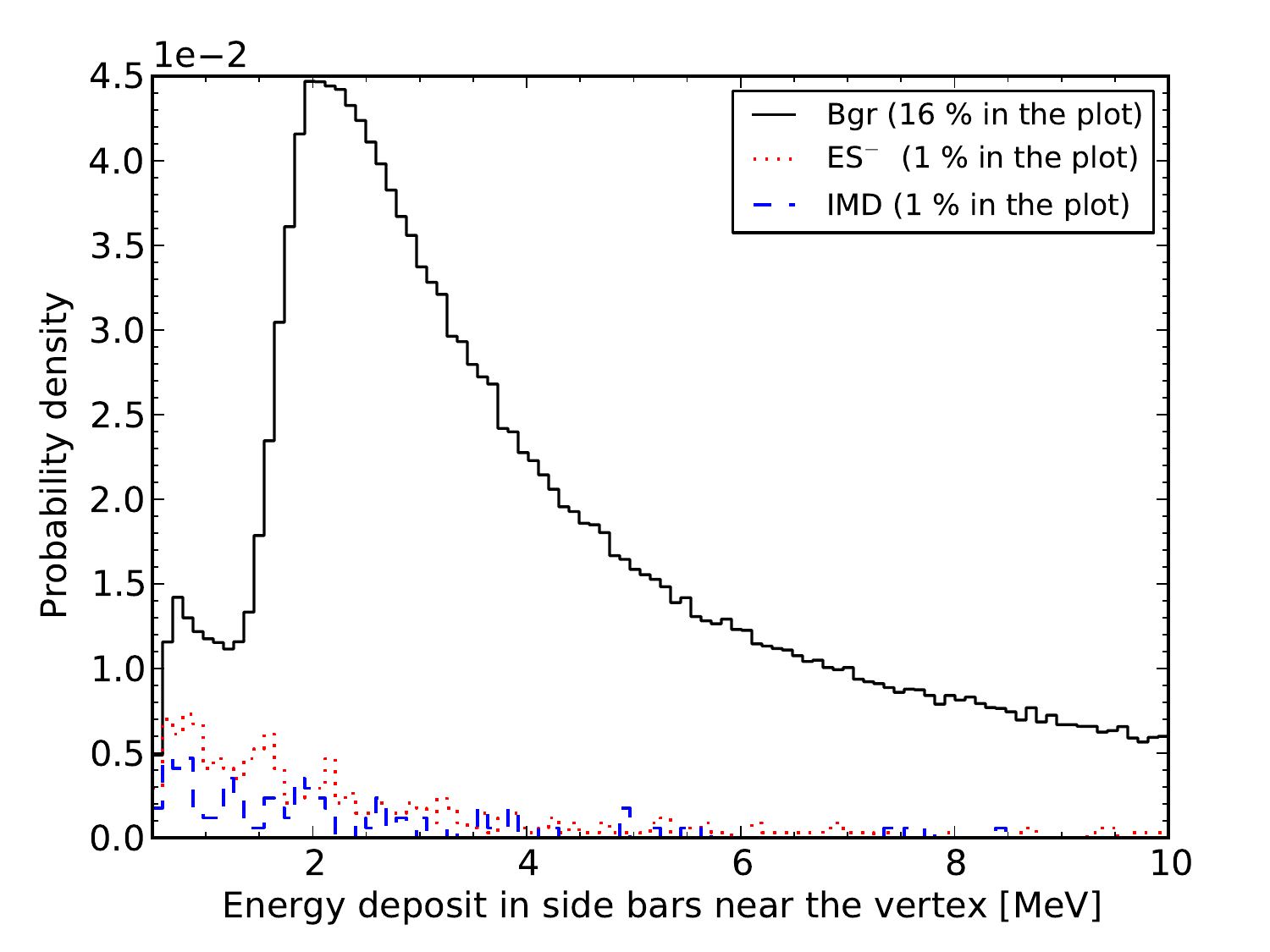}
  \caption[Distributions of energy deposit in side slabs near vertex.]%
  {Distributions of total energy deposit in bars covering air gaps adjacent to the vertex for 
  \IMD (blue), \ESm (red) and background (black) events in \Pmuon mode. The fraction of events 
  contained in the plot is indicated in the legend. All distributions are normalized to a unit 
area.}
  \label{fig:bgrSideDeposit}
\end{figure}

Both \IMD and \ES events have a property of low (consistent with single particle) energy 
deposition near the vertex. To exploit that, a cut on energy deposit in the vertex bar of 
\unit{4}{\MeV} is imposed. If vertex is in a cluster of fibers, a cut is made on its amplitude 
(the sum of fiber amplitudes). In some background events, energetic (hundreds of \MeV) charged 
hadrons escape through the 
air gaps leaving small or no deposits in the calorimetric sections. Therefore, it is required that 
there is no activity in side bars covering air gaps adjacent to the vertex. Another vertex related 
cut is the requirement that there is no energy deposits upstream of the vertex (backward deposits).
Distributions of energy deposit in the vertex bar and energy deposit in side slab for signal and 
background events are shown on \FigureRef{fig:bgrVertex} and \FigureRef{fig:bgrSideDeposit}.

\subsection{Other calorimetric selections}
\begin{figure}
  \includegraphics[width=0.49\linewidth]{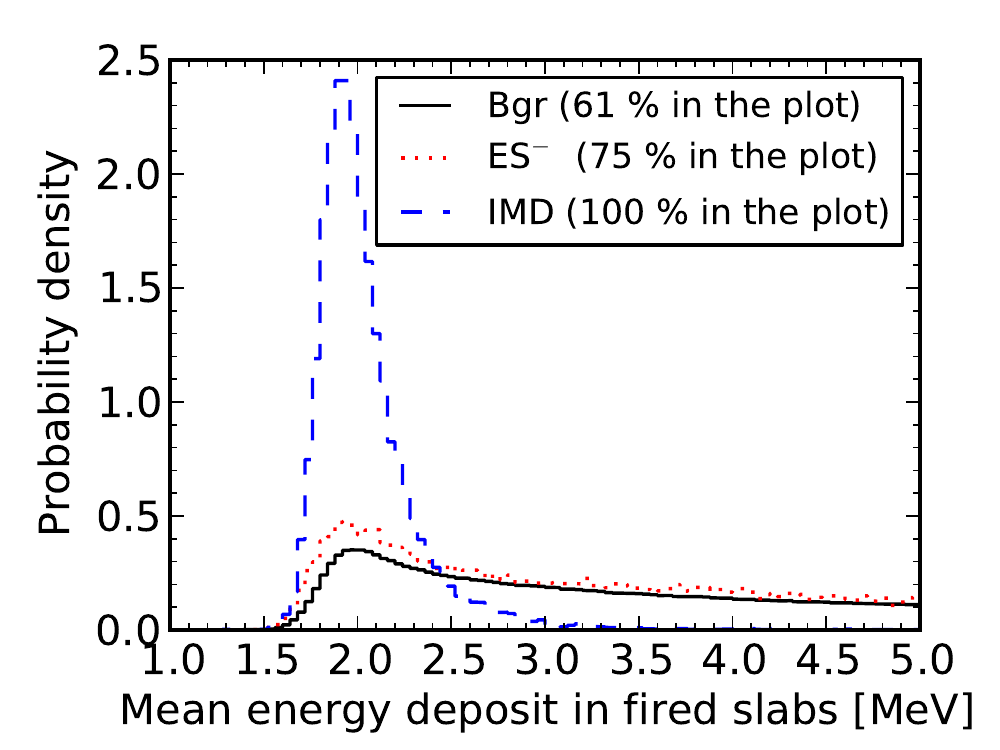}
  \includegraphics[width=0.49\linewidth]{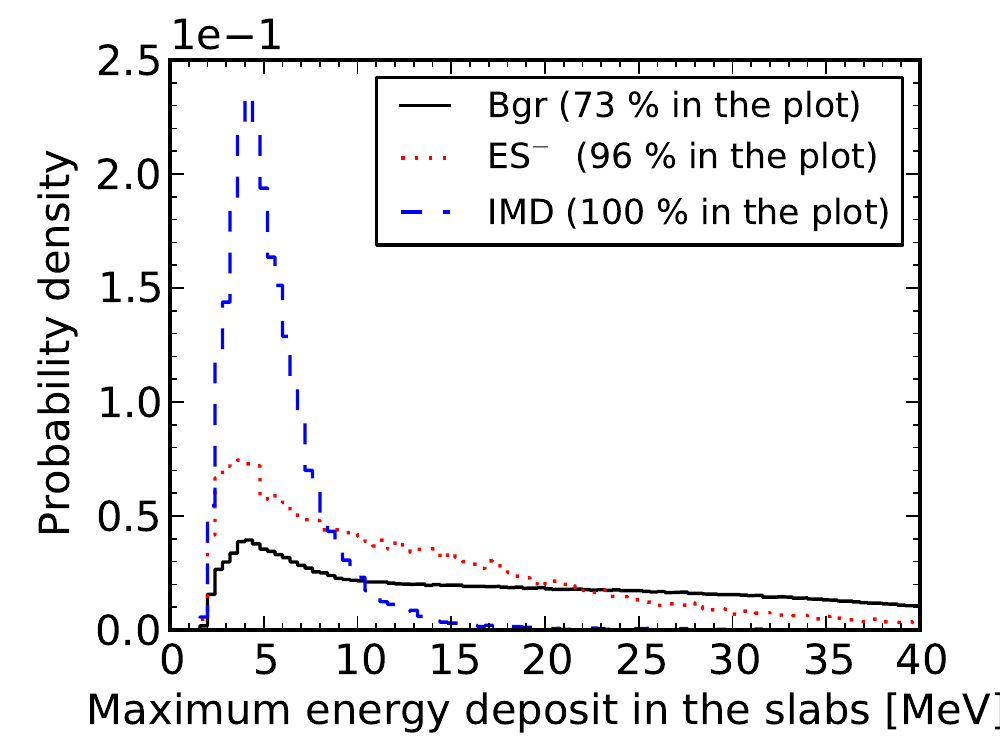}
  \caption[Distributions of $\angles{\Delta E_i}$ and $max(\Delta E_i)$.]%
  {Distributions of $\angles{\Delta E_i}$ (left) and $max(\Delta E_i)$ (right) for \IMD (blue), 
  \ESm (red) and background (black) events in \Pmuon mode. The fraction of events contained in the 
  plot is indicated in the legend. All distributions are normalized to a unit area.}
  \label{fig:bgrMeanMax}
\end{figure}

\begin{figure}
  \includegraphics[width=\largefigwidth]{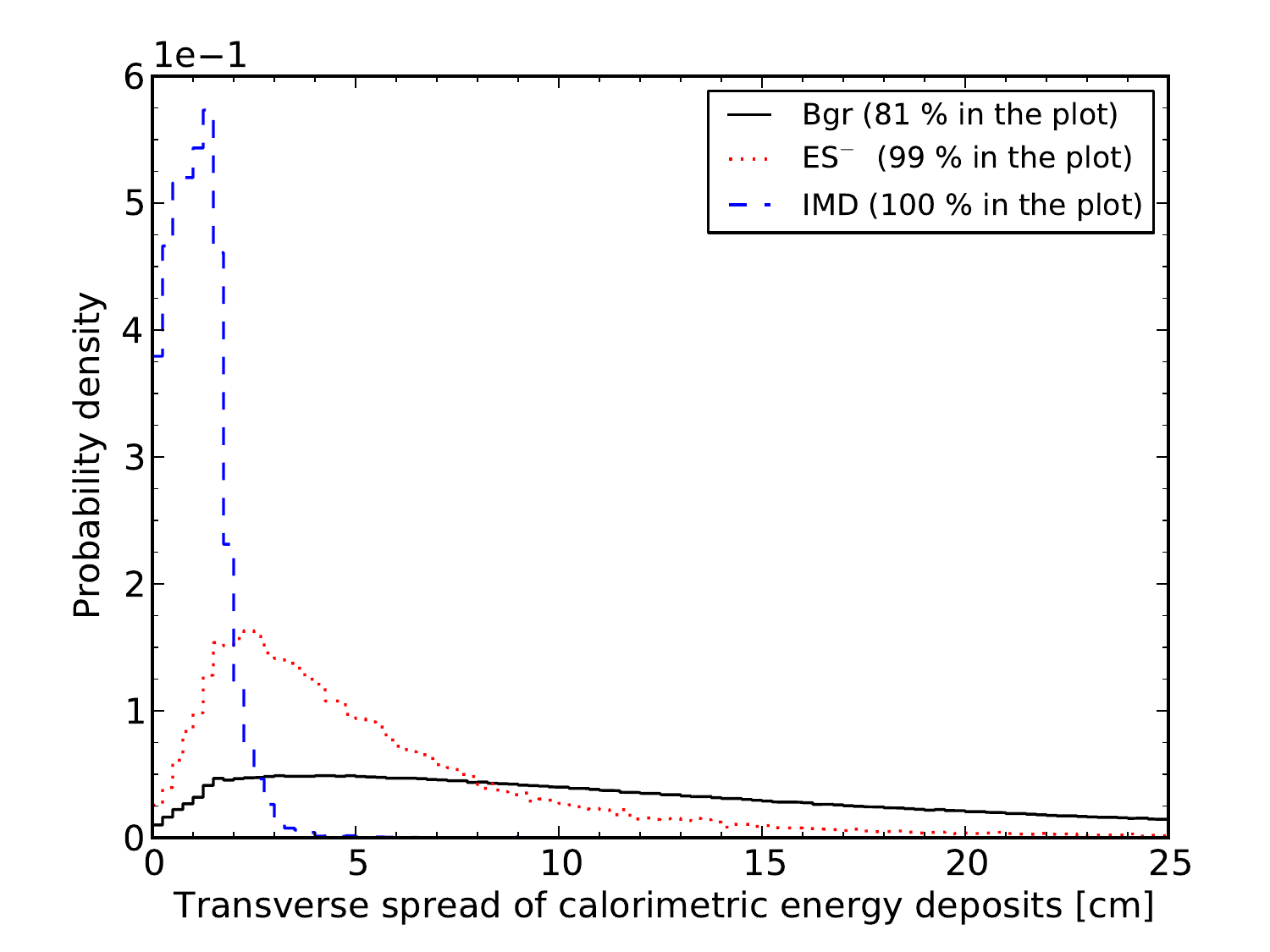}
  \caption[Distributions of $x_{dev}$ variable.]%
  {Distributions of the $x_{dev}$ variable for \IMD (blue), \ESm (red) and background (black) 
  events in \Pmuon mode. The fraction of events contained in the plot is indicated in the legend. 
  All distributions are normalized to a unit area.}
  \label{fig:bgrMeandev}
\end{figure}

To select \IMD events, one can rely on the specific properties of muon $dE/dx$. At energies in the 
range of \unit{11-25}{\GeV}, a muon is nearly a minimum ionizing particle. Therefore, the 
following cuts 
were applied when selecting \IMD events: mean of all slab deposits $\angles{\Delta E_i}$ is less 
than \unit{3}{\MeV} and the maximum $max(\Delta E_i)$ is less than \unit{12}{\MeV}. For \ES 
events, primary electron might induce shower in 
the detector, thus such cuts are not suitable. Signal and background distributions of 
$\angles{\Delta E_i}$ and $max(\Delta E_i)$ are shown on \FigureRef{fig:bgrMeanMax}.
For selecting \IMD events, the $x_{dev}$ variable is cut at \unit{2.5}{\cm}, while for \ES events 
a more relaxed cut at \unit{15}{\cm} is applied. Signal and background distribution of the 
$x_{dev}$ variable are shown on \FigureRef{fig:bgrMeandev}.

\subsection{Kinematic selections}
\begin{figure}
  \includegraphics[width=0.49\linewidth]{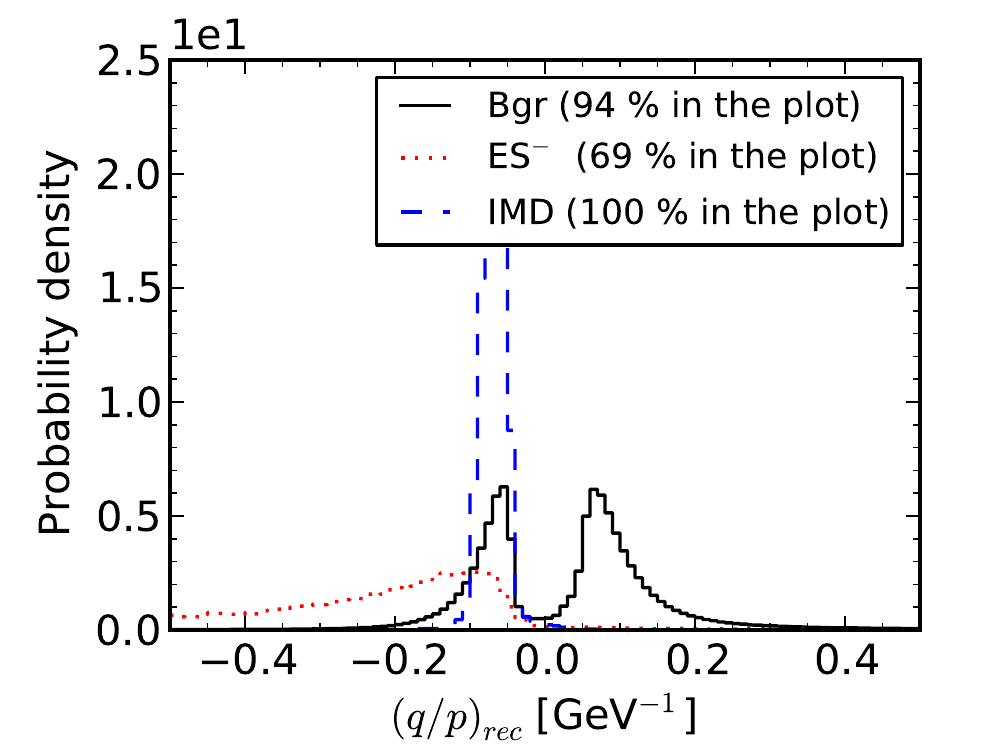}
  \includegraphics[width=0.49\linewidth]{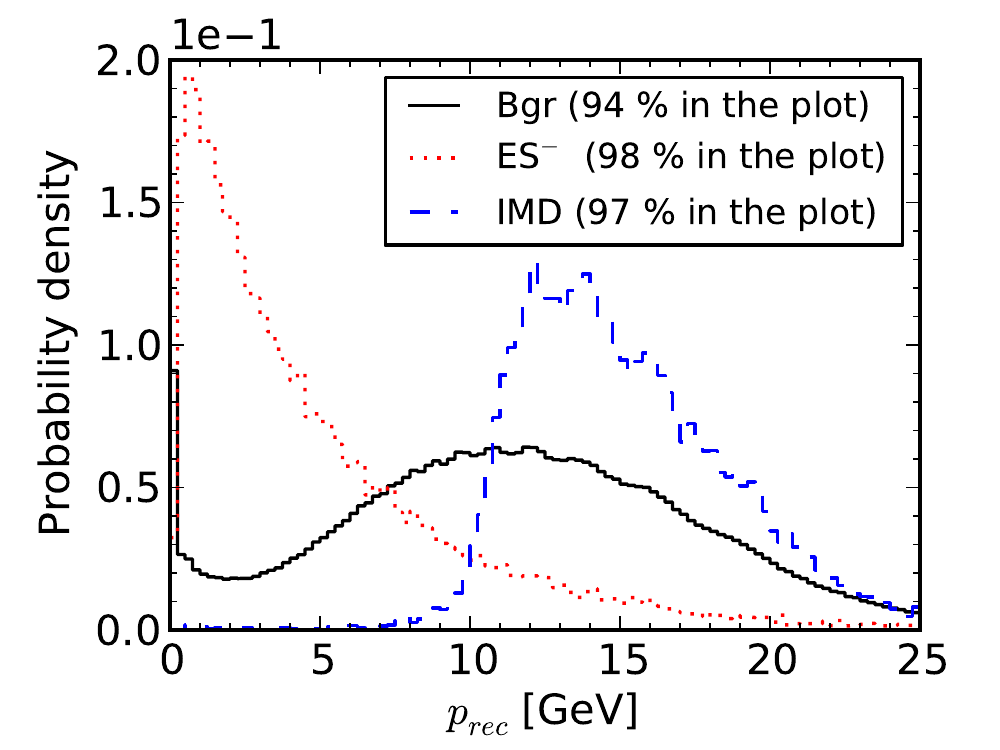}
  \caption[Distributions of primary track $q/p$ and momentum.]%
  {Distributions of primary track reconstructed $q/p$ (left) and momentum (right) for \IMD 
  (blue), \ESm (red) and background (black) events in \Pmuon mode. The fraction of events     
  contained in the plot is indicated in the legend. All distributions are normalized to a unit 
area.}
  \label{fig:bgrPrimaryQOP}
\end{figure}

As signal events always have negatively charged primary lepton (\Pmuon or \Pelectron), a cut $q/p 
< 0$ on the primary track is imposed. Quasi-elastic neutrino electron scattering has a threshold 
at $\sim \unit{11}{\GeV}$. Therefore, when selecting the \IMD sample, events with primary track 
momenta of less than \unit{10}{\GeV} are discarded. In order to avoid contamination of \IMD events 
in the \ESm sample, only events with primary track momentum less than \unit{10}{\GeV} are retained 
in this sample.%
\footnote{A full containment calorimeter enclosing the detector will be able to 
distinguish electrons from muons with high confidence. However, such additional calorimeter is 
not envisaged in the current near detector design and the author prefers not to make assumptions 
on electron/muon discriminating capabilities.}.
This is a reasonable cut, since majority of \ESm events have primary track momentum of 
less than \unit{10}{\GeV}. Distributions of primary track's reconstructed $q/p$ and momentum are 
shown on \FigureRef{fig:bgrPrimaryQOP}.

The overall efficiencies of the selection cuts are \unit{89}{\percent} for \IMD sample,
\unit{73}{\percent} for \ESm sample and \unit{87}{\percent} for \ESp sample. All cuts with their 
efficiency and purity%
\footnote{Purity is the ratio $\frac{N_{signal}}{N_{signal} + N_{background}}$.}
are summarized in \AppendixRef{app:cutsSummary}.

\section{Background subtraction}
\label{sec:bgrSubtraction}
Absolutely clean sample of signal events cannot be selected with a reasonable efficiency by 
employing selection cuts only. 
Therefore extrapolations of certain background distributions should be made in order to subtract 
background from event samples. Background subtraction is different from event selection in the 
sense that after it 
is done, one obtains a distribution of signal events and loses event by event information.

\begin{figure}
  \includegraphics[width=0.49\textwidth]{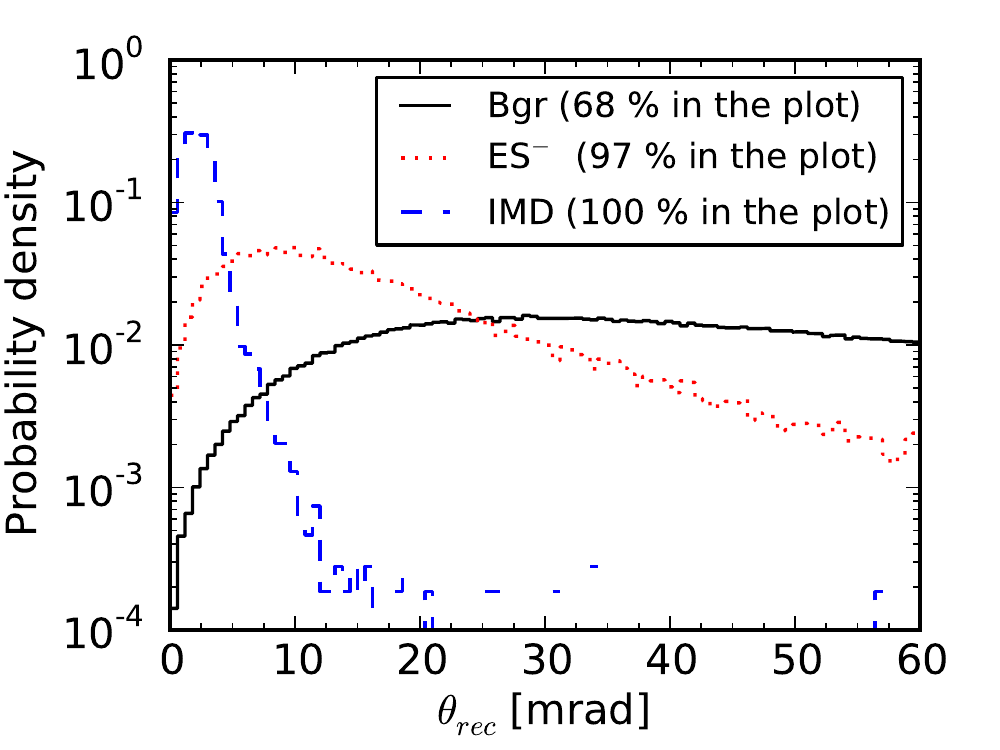}
  \includegraphics[width=0.49\textwidth]{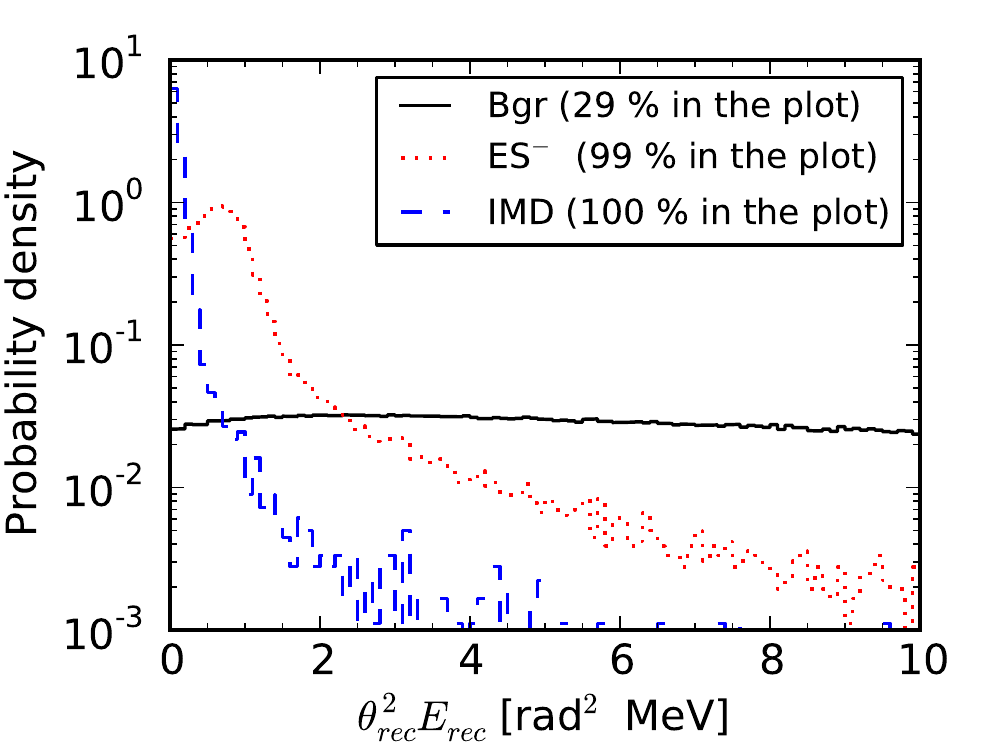}
  \caption[Distributions of primary track \scangle and \inelast.]%
  {Distributions of reconstructed primary track initial angle \scangle and \inelast variable for 
  \IMD (blue), \ESm (red) and background (black) events in \Pmuon mode. The fraction of events 
  contained in the plot is indicated in the legend. All distributions are normalized to a unit 
  area.}
  \label{fig:bgrKin}
\end{figure}

We have chosen to do background subtraction in terms of primary lepton (\Pmu or \Pe) 
kinematic variables for the following reasons:
\begin{itemize}
  \item they provide the most powerful signal/background separation criteria;
  \item background distributions can be most reliably extrapolated in terms of kinematic 
  variables;
  \item we can reliably measure primary lepton angular and momentum resolutions.
\end{itemize}
The primary lepton's scattering angle \scangle and initial momentum $p$ are measured in the 
detector. Primary muons with energy above \unit{1}{\GeV} and all primary electrons are 
ultrarelativistic. Therefore, it is justified to approximate the primary lepton's initial energy 
with $E = pc$.
For simplicity, we will perform one dimensional analysis, where histograms over a variable $x = 
f(\theta, p)$ will be used for background subtraction.
In the case of \IMD signal extraction, scattering angle \scangle and \inelast variable (see 
\SectionRef{sec:experimentalSignatures})
can be used to discriminate signal from background. In the case of \ES signal, background is well 
separated only when one expoloits \inelast variable. The distributions of \scangle and \inelast 
for \IMD signal, \ESm signal and background are shown on 
\FigureRef{fig:bgrKin}. Background distribution over \inelast variable is nearly flat. This fact 
allows for a simple parameterization of the background distribution. Our further analysis is made 
using only \inelast variable.

As a first approximation we assume that the energy distribution of neutrinos in the beam is known. 
Then, by measuring only the number of signal events, one can make absolute flux normalization.
Two methods of obtaining the number of signal events are discussed below.

\subsection{Linear fit method}
\begin{figure}
  \includegraphics[width=\largefigwidth]{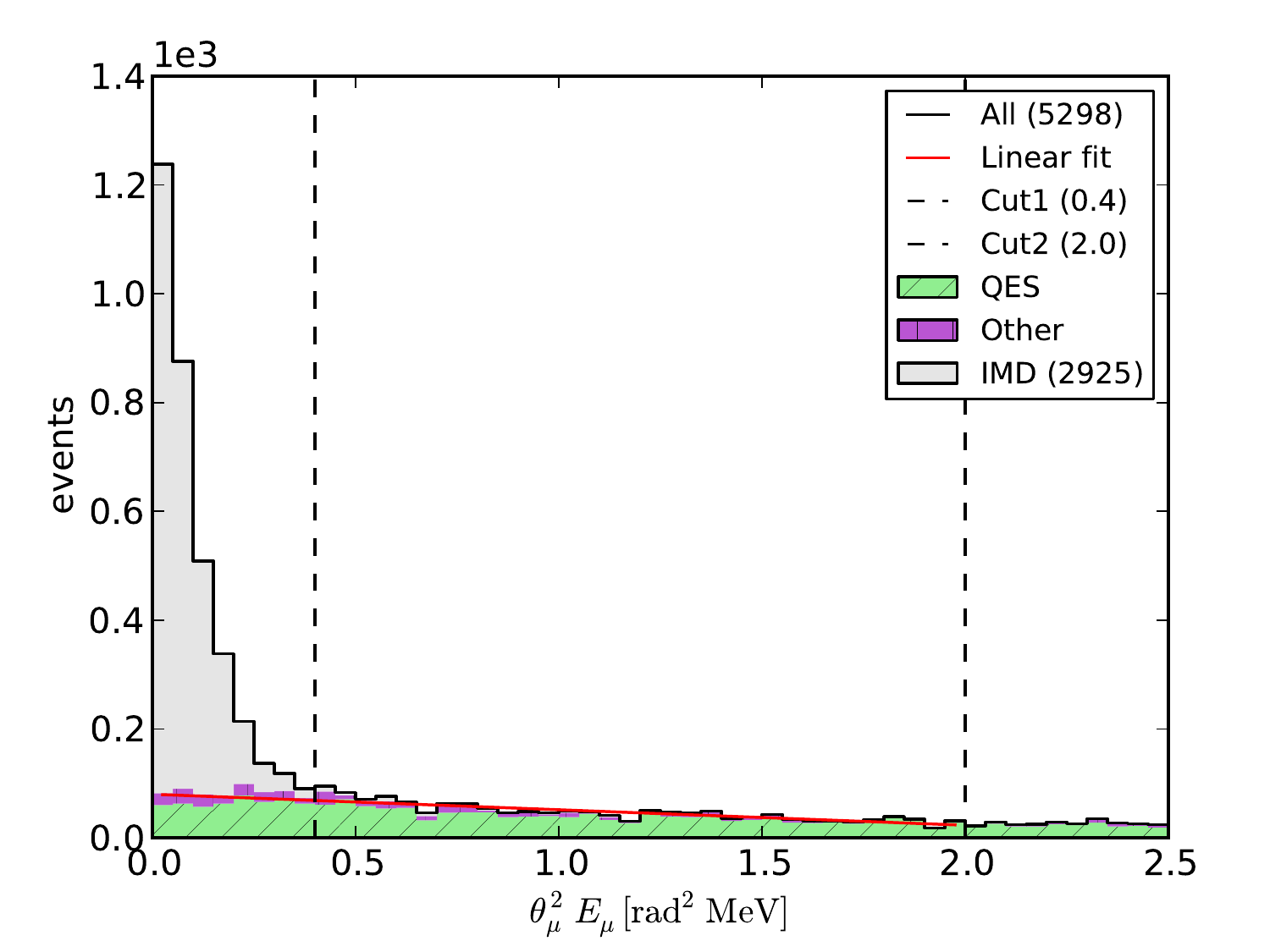}
  \caption[Distributions over \inelastMu for the \IMD sample. Linear fit method for background 
  subtraction.]%
  {Distributions over \inelastMu for the \IMD sample. The leptonic events histogram is
  filled with solid gray, the hadronic events histogram is hatched and the total spectrum is in 
  black. The two cuts bounding the fit interval are drawn with dashed line. The red line
  indicates the background extrapolation.}
  \label{fig:imdLinear}
\end{figure}

\begin{figure}
  \includegraphics[width=\largefigwidth]{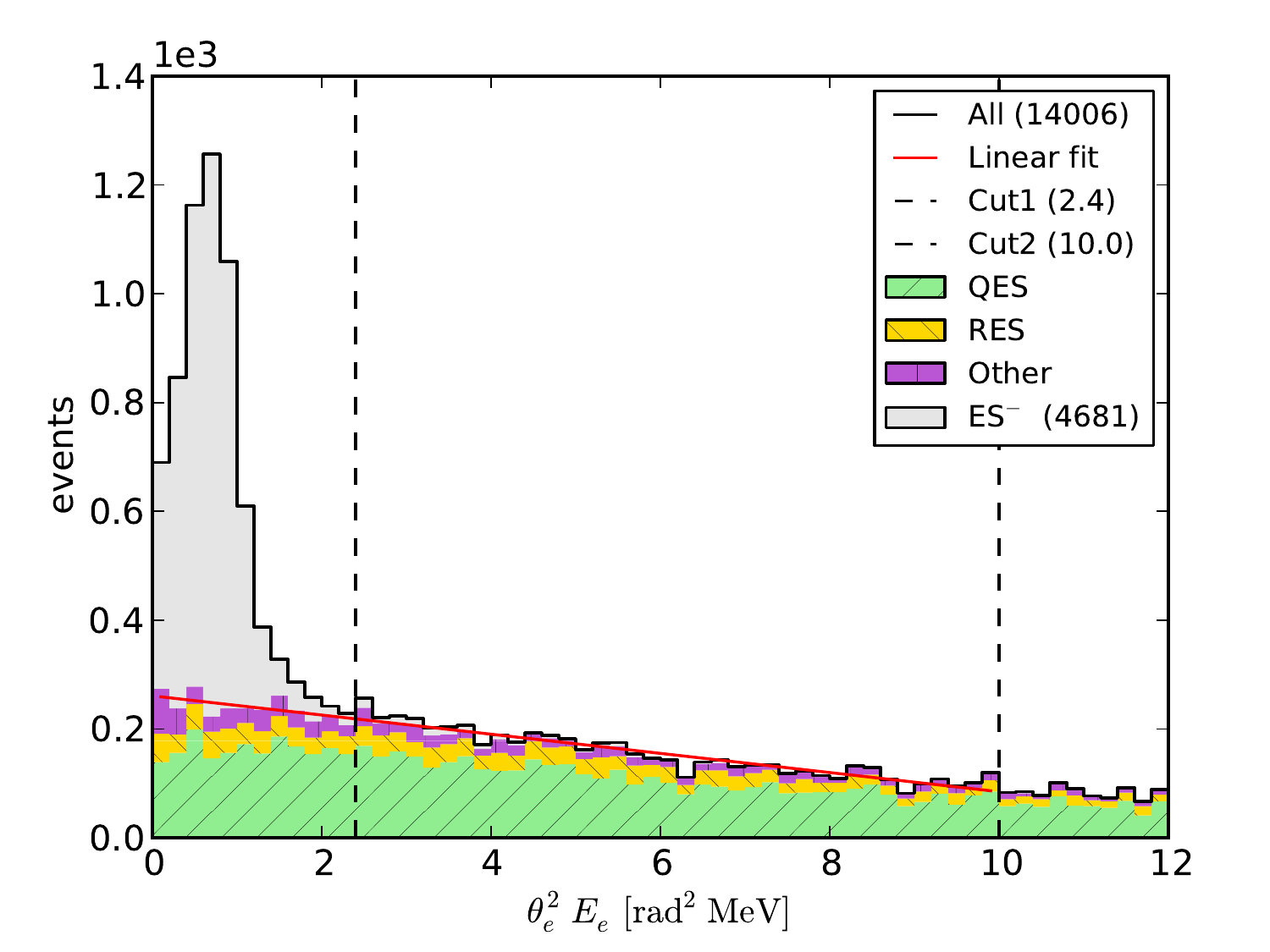}
  \caption[Distributions over \inelastE for the \ESm sample. Linear fit method for background 
  subtraction.]%
  {Distributions over \inelastE for the \ESm sample. The leptonic events histogram is
  filled with solid gray, the hadronic events histogram is hatched and the total spectrum is in 
  black. The two cuts bounding the fit interval are drawn with dashed line. The red line
  indicates the background extrapolation.}
  \label{fig:esmLinear}
\end{figure}

\begin{figure}
  \includegraphics[width=\largefigwidth]{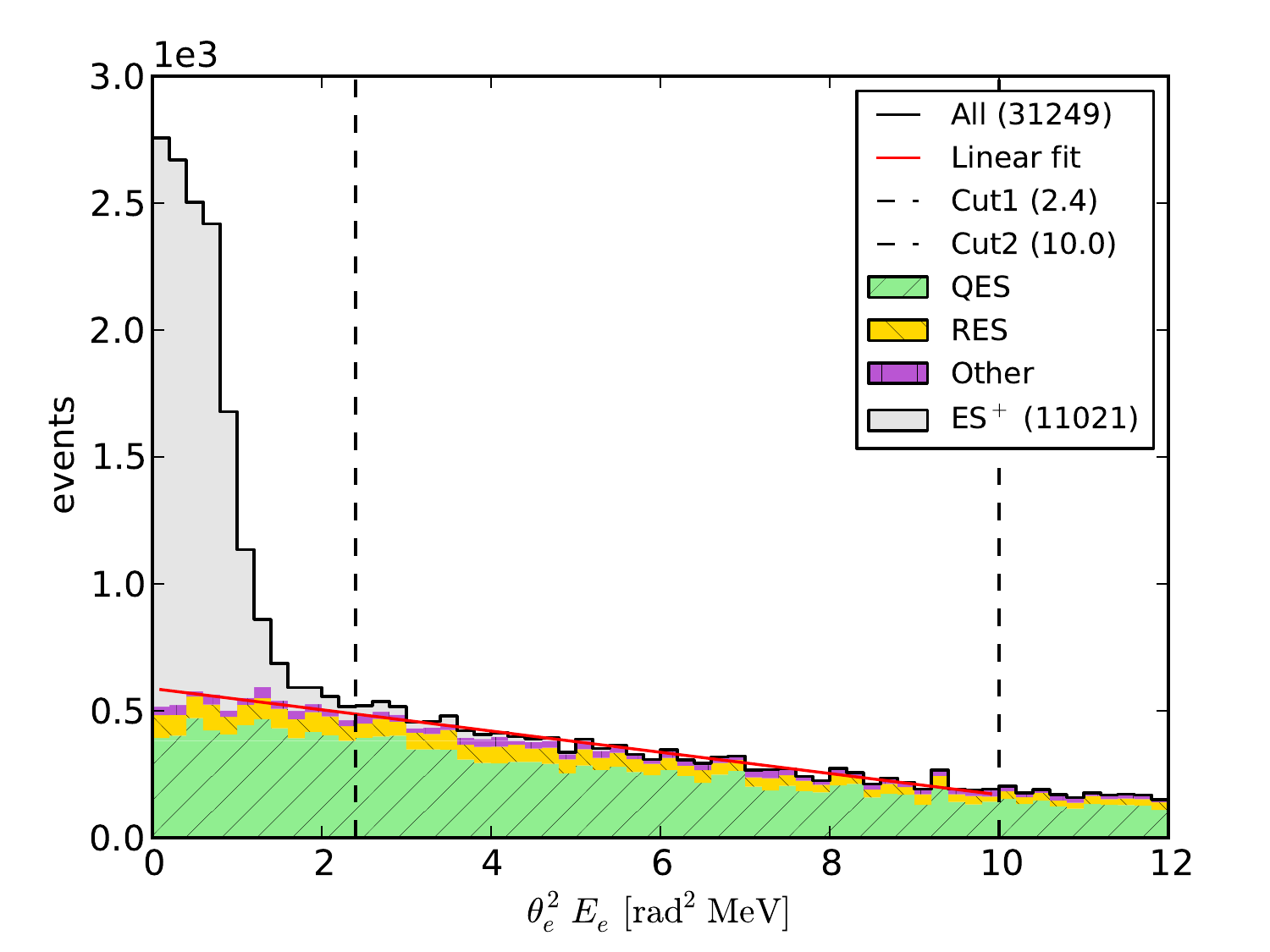}
  \caption[Distributions over \inelastE for the \ESp sample. Linear fit method for background 
subtraction.]%
  {Distributions over \inelastE for the \ESp sample. The leptonic events histogram is
  filled with solid gray, the hadronic events histogram is hatched and the total spectrum is in 
  black. The two cuts bounding the fit interval are drawn with dashed line. The red line
  indicates the background extrapolation.}
  \label{fig:espLinear}
\end{figure}

Linear fit method relies on the nearly flat shape of the respective background distribution. The 
idea is to estimate the background under the signal peak by linear extrapolation from 
signal-free region.
First, an interval on the \inelast variable is defined by the following rules:
\begin{itemize}
  \item the interval low limit is close to the signal peak;
  \item there are almost no signal events in the interval (according to MC simulation);
  \item the background is approximately linear in the interval.
\end{itemize}
Then, the histogram is fitted with a straight line in the interval. Finally, the line is 
extrapolated towards zero to estimate the number of background events under the signal peak.
The histograms over \inelast and the linear fits for the three event samples under consideration 
are shown in Figures \ref{fig:imdLinear}, \ref{fig:esmLinear} and \ref{fig:espLinear}. Comparison 
between the estimated and the true number of signal events is given on \TableRef{tab:fitResults}. 
It is seen that the true values lie within the \unit{95}{\percent} confidence intervals of the 
predictions. This means that the hypothesis that the predictions are unbiased cannot be excluded. 
The main source of systematic error related to the method is the assumption about the linearity of 
the background shape. From the fit of the \ESm and \ESp samples it is estimated that the
systematic error of the method
\begin{equation}
  \Delta_{syst} = \frac{|N_{fit}^{sig} - N_{true}^{sig}|}{N_{true}^{sig}} ,
\end{equation}
is less than \unit{1}{\percent}. However, to give conclusive estimation of the systematic error, 
one should investigate if and how various parameters of simulation and selections influence the 
background shape.

\subsection{$\mu^+$-method}
\begin{figure}
  \includegraphics[width=0.49\linewidth]{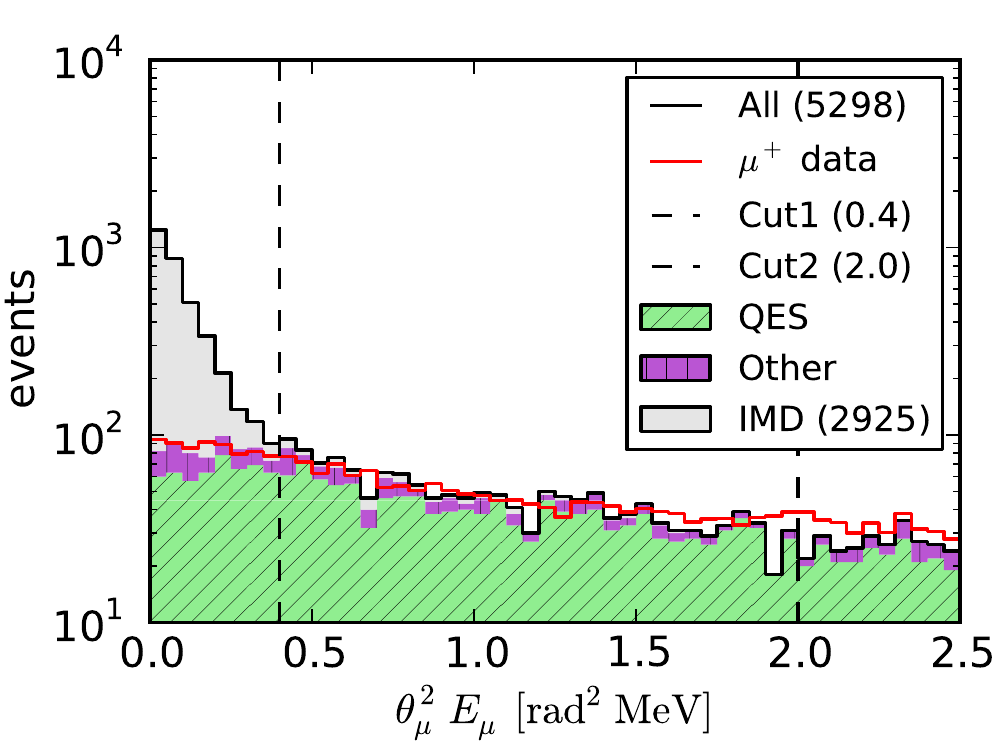}
  \includegraphics[width=0.49\linewidth]{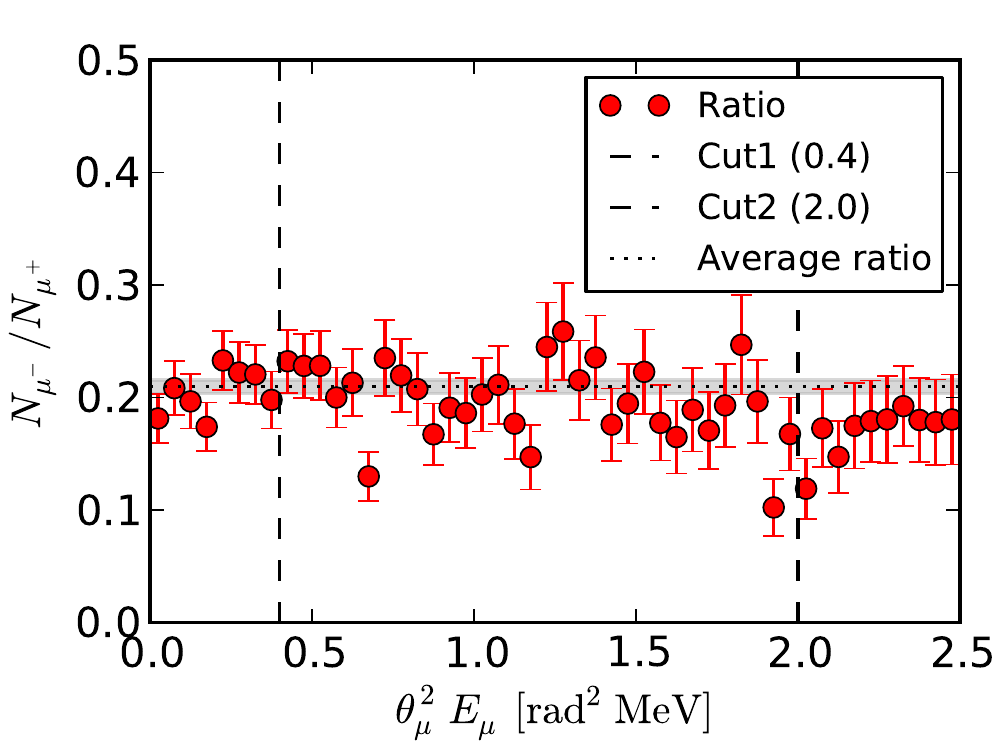}
  \caption[Distributions over \inelastMu for the \IMD sample. \APmuon-method for background 
  subtraction.]%
  {Left plot shows distributions over \inelastMu for the \IMD sample. The leptonic events 
  histogram is filled with solid gray, the hadronic events histogram is hatched and the total 
  spectrum is in black. The two cuts bounding the normalization interval $\Delta$ are drawn with 
  dashed line. The red line indicates the normalized \APmuon histogram. Right plot shows ratio of 
  the \Pmuon histogram and the \APmuon histogram over \inelastMu. The errors are statistical and 
  bars  correspond to $1 \sigma$. Horizontal dotted line indicates the normalization ratio.
  }
  \label{fig:imdMuplus}
\end{figure}

\IMD interactions are present only in the \Pmuon decay mode. The idea of the \APmuon-method is to 
estimate the background under the \IMD signal peak exploiting the distribution of positive muons 
detected in (\APnum, \Pnue)-beam. The idea was previously used for measuring \IMD process in the 
CHARM~II detector \cite{Geiregat:1990rb, Vilain:1996yf}. In the near detector, an event sample 
from the (\APnum, \Pnue)-beam events is selected with the same selection cuts as the \IMD sample. 
For consistency, 
we again consider the \inelastMu distributions. Muon antineutrino inclusive CC interactions 
$\APnum N$ have approximately twice higher cross section than $\Pnum N$ interactions. To account 
for that, the \inelastMu histogram for \APmuon should be normalized to the \inelastMu 
histogram for \Pmuon. First, the ratio of the \Pmuon background histogram and the \APmuon 
histogram is calculated - \FigureRef{fig:imdMuplus} (right). It is seen, that the ratio under the 
signal peak ($\inelastMu < \unit{0.4}{\rad^2\MeV}$) is at the same level as 
the ratio outside the signal peak. An interval outside the \IMD signal peak and with approximately 
constant ratio of \Pmuon- and \APmuon-events is defined: $\Delta = \unit{[0.4, 2.0]}{\rad^2\MeV}$. 
The normalization factor is then
\begin{equation}
  R = \frac{N_{\Pmuon}^\Delta}{N_{\APmuon}^\Delta} = 0.210 \pm 006 ,
\end{equation}
where $N_{\Pmuon}^\Delta$ ($N_{\APmuon}^\Delta$) is the number of events in the interval $\Delta$ 
for the \Pmuon (\APmuon) histogram. The \Pmuon histogram and the normalized \APmuon histogram are 
shown in \FigureRef{fig:imdMuplus} (left). Apart from normalization, there are second order 
differences between \Pmuon and \APmuon distributions \cite{Geiregat:1990rb, Vilain:1996yf}, which 
are not taken into account.

\begin{table}
  \caption[Estimated number of signal events.]%
    {Estimated number of signal events for the three event samples. The result in the last row was 
      obtained using the \APmuon background subtraction method, while the other three results were 
      obtained using linear fit background subtraction method.
      Statistics correspond to $2.3\times 10^{19}$ \Pmuon decays and $2.3\times 10^{19}$ \APmuon 
      decays, which is approximately a tenth of the nominal year.}
  \vspace{2ex}
  \begin{tabular}{lcccccc}
    Event  & Selection & Overall & Purity & All events & Signal events & Signal events \\
    sample & eff.      & eff.    &        &            &               & from fit      \\
    \midrule
    \IMD         & \unit{86}{\percent} & \unit{46}{\percent} & \unit{81}{\percent} &
      3520 & 2850 & 2926 $\pm$ 59 \\
    \ESm         & \unit{70}{\percent} & \unit{32}{\percent} & \unit{61}{\percent} &
      7355 & 4491 & 4479 $\pm$ 86 \\
    \ESp         & \unit{83}{\percent} & \unit{37}{\percent} & \unit{63}{\percent} &
      16964 & 10607 & 10512 $\pm$ 131 \\
    \midrule
    \IMD         & \unit{86}{\percent} & \unit{46}{\percent} & \unit{81}{\percent} &
     3520 & 2850 & 2831 $\pm$ 61 \\
      \end{tabular}

  \label{tab:fitResults}
\end{table}

%% file: conclusions.tex
\chapter{Conclusions}
\label{chap:conclusions}
We have proposed a conceptual design for the tracker part of the \NuFact near detector. A full 
Monte Carlo 
simulation of neutrino interactions and particle transport was implemented using standard tools.
An idealized detector geometry and simplified signal digitization was used in the MC simulation. 
Basic reconstruction algorithms were developed to extract information relevant for 
neutrino-electron scattering measurements. Selection procedures, which increase signal to 
background ratio from $\sim 10^{-4}$ initially to 
$\sim \unit{30-50}{\percent}$, are defined.
One dimensional signal extraction analysis was performed. It is demonstrated that the number of 
neutrino-electron scattering events can be measured from the \inelast variable histogram. A direct 
comparison between measured and true number of signal events shows a deviation of no more than 
\unit{1}{\percent}. It is worth noting that MC truth was not used in reconstruction and signal 
extraction. 
Therefore, the results obtained in this work are a reliable approximation to results obtainable at 
a real experiment, provided that MC simulation is close to reality.

Further developments of the methods used towards a conclusive statement about the achievable 
precision of neutrino flux measurement are needed.
A two-dimensional signal extraction procedure in the $(\theta_{\Pl}, \p_{\Pl})$ plane
will allow the extraction of lepton energy distribution of signal events. 
Subsequently, a distribution of neutrino energy should be obtained.
Development of advanced pattern recognition and fitting reconstruction algorithms is expected to 
improve on background rejection and angular and momentum resolutions.
Finally, a dedicated analysis of systematic uncertainties should be made to asses the precision of 
the neutrino flux determination.

%% file: appendices.tex

\chapter{Summary of cuts}
\label{app:cutsSummary}
Tables \ref{tab:cutsIMD}, \ref{tab:cutsESm} and \ref{tab:cutsESp} summarize the selection cuts for 
the \IMD, \ESm and \ESp signal event samples. Only the efficiencies of the selection cuts (not 
including reconstruction efficiency) are given.

\begin{table}[bht]
  \caption[Selection cuts for the \IMD sample.]%
    {Selection cuts for the \IMD sample.}
  \vspace{2ex}
  \begin{tabular}{rlrr}
    Cut & & Eff. after cut & Purity after cut \\
    \midrule
    All events in detector& & & $0.85 \times 10^{-4}$ \\
    Reconstruction selections & & & \unit{0.19}{\percent} \\
    \midrule
    Side slabs deposit    & $ = \unit{0}{\MeV}$ & \unit{99}{\percent} & \unit{0.25}{\percent} \\
    Backward deposit      & $ = \unit{0}{\MeV}$ & \unit{98}{\percent} & \unit{0.28}{\percent} \\
    Vertex deposit        & $ < \unit{4}{\MeV}$ or &  \\
    & $ < 500$ p.e. & \unit{98}{\percent} & \unit{0.47}{\percent} \\
    $\angles{\Delta E_i}$ & $ < \unit{3}{\MeV}$  & \unit{96}{\percent} & \unit{1.4}{\percent} \\
    $\max(\Delta E_i)$    & $ < \unit{12}{\MeV}$ & \unit{94}{\percent} & \unit{1.5}{\percent} \\
    $x_{dev}$             & $ < \unit{2.5}{\cm}$ & \unit{92}{\percent} & \unit{13}{\percent} \\
    $(q/p)_{rec}$         & $ < 0 $              & \unit{91}{\percent} & \unit{28}{\percent} \\
    $p_{rec}$             & $ > \unit{10}{\GeV}$ & \unit{89}{\percent} & \unit{34}{\percent} \\
  \end{tabular}
  \label{tab:cutsIMD}
\end{table}
\begin{table}[bht]
  \caption[Selection cuts for the \ESm sample.]%
  {Selection cuts for the \ESm sample.}
  \vspace{2ex}
  \begin{tabular}{rlrr}
    Cut & & Eff. after cut & Purity after cut \\
    \midrule
    All events in detector& & & $2.0 \times 10^{-4}$ \\
    Reconstruction selections & & & \unit{0.37}{\percent} \\
    \midrule
    Side slabs deposit    & $ = \unit{0}{\MeV}$ & \unit{98}{\percent} & \unit{0.48}{\percent} \\
    Backward deposit      & $ = \unit{0}{\MeV}$ & \unit{95}{\percent} & \unit{0.52}{\percent} \\
    Vertex deposit        & $ < \unit{4}{\MeV}$ or &  \\
    & $ < 500$ p.e.          & \unit{94}{\percent} & \unit{0.89}{\percent} \\
    $x_{dev}$             & $ < \unit{15}{\cm}$  & \unit{89}{\percent} & \unit{1.5}{\percent} \\
    $(q/p)_{rec}$         & $ < 0 $              & \unit{86}{\percent} & \unit{7.2}{\percent} \\
    $p_{rec}$             & $ < \unit{10}{\GeV}$ & \unit{73}{\percent} & \unit{23}{\percent} \\
  \end{tabular}
  \label{tab:cutsESm}
\end{table}
\begin{table}[bht]
  \caption[Selection cuts for the \ESp sample.]%
  {Selection cuts for the \ESp sample.}
  \vspace{2ex}
  \begin{tabular}{rlrr}
    Cut & & Eff. after cut & Purity after cut \\
    \midrule
    All events in detector& & & $4.1 \times 10^{-4}$ \\
    Reconstruction selections & & & \unit{0.67}{\percent} \\
    \midrule
    Side slabs deposit    & $ = \unit{0}{\MeV}$ & \unit{98}{\percent} & \unit{0.83}{\percent} \\
    Backward deposit      & $ = \unit{0}{\MeV}$ & \unit{95}{\percent} & \unit{0.89}{\percent} \\
    Vertex deposit        & $ < \unit{4}{\MeV}$ or &  \\
    & $ < 500$ p.e.          & \unit{94}{\percent} & \unit{1.4}{\percent} \\
    $x_{dev}$             & $ < \unit{15}{\cm}$  & \unit{90}{\percent} & \unit{2.2}{\percent} \\
    $(q/p)_{rec}$         & $ < 0 $              & \unit{87}{\percent} & \unit{24}{\percent} \\
  \end{tabular}
  \label{tab:cutsESp}
\end{table}

%% file: backmatter.tex
\begin{acknowledgements}
I would like to express my gratitude for my supervisor Professor Roumen Tsenov who has 
continuously advised and supported me throughout my studies and thesiswork. I would also like to 
thank Dr. Yordan Karadzhov for introducing me to the bits and pieces of high energy physics 
simulations. I am thankful to all colleagues from the neutrino physics group for the warm and 
fruitful environment they have created.
\end{acknowledgements}

\bibliographystyle{utphys}
\bibliography{mythesis}
